\documentclass[preprints,review,accept,moreauthors,pdftex]{Definitions/mdpi} 

\firstpage{1} 
\pubvolume{1}
\issuenum{1}
\articlenumber{0}
\pubyear{2021}
\copyrightyear{2020}
\datereceived{} 
\dateaccepted{} 
\datepublished{} 
\hreflink{https://doi.org/} 

\Title{GRB Prompt Emission: Observed Correlations and Their Interpretations}

\TitleCitation{GRB Prompt Emission: Observed Correlations and Their Interpretations}

\Author{Tyler Parsotan $^{1,2,3,\dagger,\ddagger}$\orcidA{} and Hirotaka Ito $^{4,5, \ddagger}$\orcidB{}}

\AuthorNames{Tyler Parsotan and Hirotaka Ito}

\AuthorCitation{Parsotan, T.; Ito, H.}

\address{%
$^{1}$ \quad Center for Space Science and Technology, University of Maryland Baltimore County, 1000 Hilltop Circle, Baltimore, MD 21250, USA.; parsotat@umbc.edu\\
$^{2}$ \quad Astrophysics Science Division, NASA Goddard Space Flight Center,Greenbelt, MD 20771, USA.\\
$^{3}$ \quad Center for Research and Exploration in Space Science and Technology, NASA/GSFC, Greenbelt, Maryland 20771, USA\\
$^{4}$ \quad Astrophysical Big Bang Laboratory, RIKEN, Saitama 351-0198, Japan; hirotaka.ito@riken.jp\\
$^{5}$ \quad Interdisciplinary Theoretical \& Mathematical Science Program (iTHEMS), RIKEN, Saitama 351-0198, Japan
}

\corres{Correspondence: parsotat@umbc.edu}

\secondnote{These authors contributed equally to this work.}

\abstract{The prompt emission of Gamma Ray Bursts (GRBs) is still an outstanding question in the study of these cataclysmic events. Part of what makes GRBs difficult to study is how unique each event seems to be. However, aggregating many GRB observations and analyzing the population allows us to obtain a better understanding of the emission mechanism that produces the observed prompt emission. In this review, we outline some of the most prevalent correlations that have emerged from GRB prompt emission observations and how these correlations are interpreted in relation to GRB physical properties and prompt emission models.}


\begin{document}

\section{Introduction}

Gamma Ray Bursts (GRBs) are the most energetic events in the universe releasing up to $10^{54}$ ergs over the course of a given event. GRBs were initially detected in the 1960's by the Vela satellite as pulses of gamma ray emission. These pulses were  determined to be of extraterrestrial origin \cite{first_grbs} and eventually of cosmological origin \citep{Reichart_1998}. 

\subsection{  Observations}
Since the 1960's a number of satellites have been launched with one of their main science goals being understanding GRBs. These satellites include: 
\begin{itemize}
    \item BeppoSAX \cite{bepposax},
    \item the Compton Gamma Ray Observatory (CGRO), with the Burst and Transient Source Experiment (BATSE) on board \citep{Fishman_BATSE},
    \item the High Energy Transient Explorer (HETE-2) with the Wide-field X-ray Monitor \cite{hete},
    \item {  the Wind Spacecraft with the Konus Wind GRB Experiment on board \cite{konus_wind}},
    \item the INTEGRAL satellite with the IBIS (Imager on-Board the INTEGRAL Satellite) instrument \citep{Ubertini_IBIS}, 
    \item the Neil Gehrels Swift Observatory \citep{Gehrels_Swift}, with the Burst Alert Telescope instrument \citep{Barthelmy_BAT}, 
    \item and the Fermi Gamma-ray Space Telescope with the Gamma Ray Burst Monitor (GBM) instrument onboard \cite{Case_GBM}.
\end{itemize}

These satellites have detected numerous GRBs over their lifetimes, providing rich datasets that have contributed to our current state of knowledge concerning GRBs. {  Both Konus-Wind and BATSE data showed} that GRBs are isotropically distributed in the sky, supporting the theory that GRBs were located at cosmological distances \citep{Fishman_BATSE, konus_isotropic_grb} before the first redshift measurement was obtained. Additionally, the BATSE dataset hinted at two distinct populations of GRBs, based on the duration of observed GRBs \citep{BATSE_catalog, kouveliotou1993identification}. This result has been reinforced by modern datasets and through many other lines of evidence. The two populations of GRBs are known as Short GRBs (SGRBs), which typically have durations $\lesssim 2$ s, and Long GRBs (LGRBs) which typically have durations $\gtrsim 2$ s. SGRBs have been associated with the merger of a Neutron star and a Black hole \cite{GW_NS_merger, grb_NS_merger_connection} and are also theorized to be caused by the merger of a pair of Neutron stars. On the other hand, LGRBs have been associated with core collapse supernovae from massive stars \cite{grb_sn_connection, hjorth2003_LGRB_SNe, woosley2006supernova_connection}{  \footnote{It is important to note that this classification based on just the duration of GRBs leaves room for ambiguity. The distinction between these two classes can be improved by including more than one classification parameter, for example plotting the GRBs on a duration-spectral hardness plane \citep{zhang2012revisiting}. The problem of classifying GRBs properly is still an active topic of research.}}. There have been additional subclasses of GRBs that have been proposed recently such as SGRBs with extended emission \citep{Norris_SGRB_EE, Gehrels_SGRB_EE} and ultra long GRBs \citep{Boer_ULGRB, Levan_ULGRB}, which are still active topics of research.

Focusing on ``traditional'' LGRBs and SGRBs, each type exhibits similar phases of emission. The initial pulse of high energy X-rays and gamma-rays that is detected within the first few tens of seconds of the event is known as the prompt emission. This phase of GRBs has also been detected in optical wavelengths (see Table 1 presented in \cite{parsotan_optical}) and has a number of polarimetry measurements associated with it (see Table 1 presented in \cite{Gill_Polarization}). The second phase of typical GRB emission is known as the afterglow. This phase is detected across the electromagnetic spectrum and lasts for timescales significantly longer than the prompt emission. As a result, the followup of GRB prompt emission detections, to conduct measurements of the afterglow phase, has contributed greatly to our understanding of GRBs. In this review we focus on the prompt emission phase of the GRB and correlations that have been discovered with respect to this phase of emission. Due to the longevity of afterglow emission, some of the correlations include GRB parameters that are derived from afterglow observations.

After decades of observing the prompt emission of GRBs, the radiation mechanism that produces this emission is still not well understood. There are a number of reviews focused on both the theory and observations of GRB prompt emission (see \cite{peer_review, zhang_review, meszaros_review, beloborodov_review, DAVANZO_review, KUMAR_review, Meszaros_review19}) which we briefly summarize here. The light curves of GRB prompt emission can be erratic and variable, with each light curve seeming to be unique (see e.g. Figure 6 in \cite{FERMI}). Typically, the pulses in GRB light curves seem to follow Fast Rise Exponential Decay (FRED) profiles \citep{norris1996_GRB_FRED_pulses, norris_fred}.  The spectra of GRB are typically non-thermal with typical low energy photon indices $\alpha \sim -1$, high energy photon indices $\beta \sim -2.4$, and spectral peak energies for $\nu$F$_\nu$ spectra of $E_\mathrm{pk} \sim 200$ keV \citep{FERMI}. There have been a number of detections of thermal components in GRB spectra \citep{Ryde_thermal, Peer_thermal, Guiriec_thermal, HIC_guiriec, Axelsson_thermal}, however these seem to be in the minority as of now.

\subsection{  Models}
There are two major classes of models used to describe the mechanism behind GRB prompt emission. The first class of models is based on synchrotron emission produced by magnetic fields within the GRB jet. These models consider both ordered and disordered magnetic fields within the GRB jet. These magnetic fields are typically constrained to shells of the GRB jet that have been launched by a central engine \cite{SSM_REES_MES,ICMART_Zhang_2010, toma2008statistical_GRB_pol}. Each shell moves with a different Lorentz factor with respect to the other shells which results in collisions between shells at large distances from the central engine ($\sim 10^{14}-10^{16}$ cm \cite{ICMART_Zhang_2010}) where the optical depth is below unity. These collisions excite electrons within the shells which then emit non-thermal synchrotron emission. The strength of the magnetic field plays an important role in the synchrotron emission as well as the time scales over which the electrons cool \citep{daigne2011marginally_fastcooling_synch}. The model that describes shells of material colliding within the jet has historically been known as the internal shock model although that name is not as widely used since there are a number of models that use the basic premise outlined previously. The general subclass of internal shock models were contrasted with the subclass of external shock models, where the GRB was thought to be produced by the jet colliding with the external medium \cite{meszaros_review}. The external shock model however, has fallen out of favor due to its inconsistency with observations \cite{ramirez-ruiz_central_engine, epk_fluence_relation}. 

The other major class of models used to describe GRB prompt emission considers the photospheres of jets, and thus is referred to as the photospheric model. This model describes thermal radiation that originates deep in the GRB jet at large optical depths. As the jet propagates it eventually becomes transparent to this radiation, allowing the radiation to freely travel towards the observer without being modified significantly by the jet when the optical depth is $\sim 1$ \citep{REES_MES_dissipative_photosphere, lazzati_photopshere, Thompson_photosphere, lumi_var_meszaros_photosphere}. This model was initially thought to produce only thermal spectra however, it has been shown to be able to produce non-thermal spectra due to a number of effects including high latitude emission \citep{Peer_multicolor_bb, parsotan_var}, subphotospheric dissipation \citep{Atul, bhattacharya2020_dissipation}, and the property of the photosphere being a volume of space (instead of a static surface) where photons can be upscattered through sparse interactions with matter in the jet\footnote{This concept has been called the photospheric region \cite{parsotan_mcrat} or the fuzzy photosphere \cite{Peer_fuzzy_photosphere, Beloborodov_fuzzy_photosphere}, in contrast with a photospheric surface.} \citep{ito_stratified_jets, parsotan_mcrat, Peer_fuzzy_photosphere, Beloborodov_fuzzy_photosphere}.

The correlations that emerge from observations of GRB prompt detections play an important role in constraining these models. These correlations provide information on what physical properties GRB jets have and allow us to infer how these properties may affect the radiation mechanism at play in these relativistic explosions. While we focus primarily on the physical interpretations of the correlations presented here in the context of prompt emission models, these correlations are also important for using GRBs as cosmological probes. In fact, this motivation led to the discovery of many correlations discussed in this review. For a comprehensive review in this context, we refer the reader to \cite{Dainotti_review}.

This review is organized as follows: Section \ref{quantities} provides an overview of various physical quantities, both observational and theoretical, that are necessary to understand the various correlations that are presented here. The correlations and their physical interpretations are presented in Section \ref{correlations}. Finally, in Section \ref{conclusion} we summarize the current status of prompt emission correlations and the future of these important tools to understanding the prompt emission of GRBs.

\section{Relevant Physical Quantities} \label{quantities}
\subsection{  Fundamental Quantities of Energy and Spectra}
A number of physical quantities are used to understand GRB detections. When a GRB is detected, the photons interact with the detector which has some effective area to the photons for a range of energies. These photons are then tagged with time, energy, and location information which can be used to construct GRB spectra and gain insight into the energetics of these events.

GRB spectra play an important role in understanding the prompt emission mechanism. GRB spectra can be fit using a number of different models but here we highlight two of the most commonly used models. These are:
\begin{itemize}
    \item[i)] the Band spectrum \cite{Band} which is defined as
    \begin{equation}
        N(E)= A
    \begin{cases} 
      E^\alpha e^{-\frac{E}{E_o}} & E\leq (\alpha-\beta)E_o \\
      [(\alpha-\beta)E_o]^{\alpha-\beta}E^\beta e^{\beta-\alpha} & E > (\alpha-\beta)E_o 
   \end{cases}
    \end{equation}
    
    \item[ii)] the Cutoff powerlaw spectrum \cite{FERMI} (also called the comptonized spectrum\footnote{This name is misleading as Comptonization can produce many types of spectra while cutoff power laws can
    be produced for a number of emission processes.}) which is defined as
    \begin{equation}
        N(E)=AE^\alpha e^{-\frac{E}{E_o}}
    \end{equation}
\end{itemize}
Here, $A$ is some normalization parameter, $\alpha$ is the low energy photon index, $\beta$ is the high energy photon index, and $E_o$ is the break energy between the low energy and  high energy power law slopes in the case of the Band spectrum, or the low energy slope and the exponential decay in the case of the cutoff powerlaw. Both models are in units of photon flux per unit energy (typically ph s$^{-1}$ cm$^{-2}$ keV$^{-1}$). The spectral peak energy of the $E^2N(E)$ spectrum for both models is located at $E_\mathrm{pk}=(2+\alpha)E_o$.


There are different observer energy-related quantities that can be constructed from a GRB observation. These are:
\begin{itemize}
    \item[i] Fluence, $S$, which has units of erg cm$^{-2}$, is the total energy emitted by the GRB as measured by a detector with some effective area within some energy band
    \item[ii] Flux, $F$, which has units erg s$^{-1}$ cm$^{-2}$, is the energy emitted by the GRB within some time period, as measured by a detector with some effective area
\end{itemize}
We obtain the average flux, $F_{\rm tot}$, in the limit that the flux is calculated over the time period in which we receive 100\% of the emission from a GRB. 

Observed GRB spectra can be produced within a given time bin and then integrated over energy in order to calculate flux within a given time bin. Alternatively, the spectra can be fit with the aforementioned models and then the models can be integrated over a larger energy range than what a given instrument might allow, providing a larger effective bandpass for calculating the flux and fluence. This is known as the k-correction which additionally allows spectra to be corrected for cosmological redshifts \cite{bloom2001prompt}. Time integrated (over the duration of the GRB) spectra are used to calculate $S$ while time-resolved spectra are used to calculate $F$. Typically, the same spectrum that is used to calculate $S$ or $F$ is used to determine the spectral $E_\mathrm{pk}$ (see e.g. the Amati relation in Section \ref{sec:amati_yonetoku}, however also see the Yonetoku relation where the time-integrated spectrum is used to calculate $E_\mathrm{pk}$ but the peak luminosity in a 1 s bin is also used in the correlation.)

The luminosity, $L$, and emitted energy, $E$, are typically calculated based on the assumption that the GRB emits its energy isotropically, within a solid angle of $4\pi$. These quantities are typically calculated in the rest frame of the GRB. The isotropic energy is calculated as 
\begin{equation}
    E=4\pi D_L^2 S
\end{equation}
where $D_L$ is the luminosity distance which depends on various cosmological parameters. Similarly, the luminosity is calculated as
\begin{equation}
    L=4\pi D_L^2 F 
\end{equation}
Different specifications of the energy range and time bins that go into the calculation of $S$ or $F$ gives rise to the different types of luminosities and energies that are quoted in the literature:
\begin{itemize}
    \item[i)] $F_{\rm tot}$ $\xrightarrow{}$ $L_{\rm iso}$ 
    \item[ii)] $F_{\rm pk}$ $\xrightarrow{}$ $L_{\rm pk}$ 
    \item[iii)] $S_{\rm tot}$ $\xrightarrow{}$ $E_{\rm iso}$ 
\end{itemize}
To give some examples, $S_{\rm tot}$ can be the (time-integrated) total energy emitted by the GRB in the energy range of 1-$10^4$ keV (see e.g. the Amati relation \cite{Amati}), the $F_{\rm pk}$ can be the peak flux measured from 30-$10^4$ keV in the time bin of interest (see e.g. the peak 1 s time bin used for the Yonetoku relation \cite{Yonetoku}), and $F_{\rm tot}$ can be the average flux measured in the energy range of 1-$10^4$ keV over the duration of the GRB. The aforementioned energy ranges are typically in the rest frame of the GRB, which makes $L$ and $E$ rest frame values. These quantities also include a k-correction since the energy ranges extend much further than a single instrument's energy range. 

\subsection{  Light Curve Morphology Quantities}
The morphology of the light curve and the time scales associated with them are also important quantities that are measured. The variability of the light curve, $V$, is typically measured by calculating a smoothed light curve and taking the difference between the smoothed light curve and the original light curve. This is done for multiple time intervals and then added over the duration of the GRB and normalized appropriately \cite{lumi_var_fenimore, lumi_var_guidorzi}. The light curve can also be characterized by its rise time, $\tau_\mathrm{RT}$, which is the time for the light curve to reach it's maximum flux \cite{lumi_rt_schaefer}. Another timescale that is important for our discussion of GRB correlations is the lag time, $\tau_\mathrm{lag}$, between photons of various energy ranges. This quantity is calculated by identifying the peak light curve times in specific instrument energy channels, which corresponds to specific photon energy ranges, and calculating the difference between them.

\subsection{  Transformation of Physical Quantities}
The spectral energies and the times measured in GRBs are cosmologically shifted. As a result, if we know the redshift $z$ of the GRB, we can transform these quantities to the rest frame of the burst (which is the same as the rest frame of the central engine). An observed energy of $E_{\rm obs}$ has a rest frame energy $E_{\rm rest}=E_{\rm obs}(1+z)$. Observed times, $T_{\rm obs}$, are transformed to rest frame times as $T_{\rm rest}=T_{\rm obs}/(1+z)$. 

The quantities described here can be defined in the observer frame or the rest frame of the GRB which can quickly become difficult to track for a newcomer to the field. We highlight any quantities defined in the observer frame with a tilde above the quantity, for example a lag measured in the observer frame will be denoted $\tilde{\tau}$. A lack of a tilde associated with the variable means that the represented quantity is defined in the rest frame of the GRB. 

\subsection{  GRB Jet Model Parameters}
GRB outflows are typically thought to be moving relativistically as a collimated outflow, or jet \cite{lazzati2018_GRB170817_afterglow, gottlieb2021structure}. These jets move with Lorentz factors, $\Gamma \gtrsim 100$ \cite{Fenimore_grb_MeV_gamma_limit}, where the bulk Lorentz factor is calculated as  $\Gamma=1/\sqrt{1-\beta^2}$. Here $\beta$ is the speed of the jet normalized by the speed of light. The Lorentz factor plays an important role as it defines the region of the jet from which the observer receives radiation from, namely that region is a size that is $\sim 1/\Gamma$ \cite{Rybiki_Lightman}. 

Another quantity of interest is the half-opening angle of the jet, $\theta_\mathrm{j}$. This quantity can help place constraints on the size and the launching mechanism of GRB jets. This quantity is difficult to measure directly but it can be done indirectly by measuring the break time in the light curve of GRB afterglows \cite{rhoads1999_afterglowdynamics, Goldstein_jet_parameters_estimate} and assuming that the medium that the jet is colliding into is either homogeneous \cite{piran1999jets} or wind-like \cite{chevalier2000wind}. With  $\theta_\mathrm{j}$, we can now calculate the actual energy emitted by a given GRB without the assumption that it emits isotropically. The non-isotropic emitted energy is calculated as
\begin{equation}
    E_\gamma=(1-\cos \theta_\mathrm{j}) E_{\rm iso}
\end{equation}

GRB jets have traditionally been modelled as a top hat jet, where the Lorentz factor is a step function out to some $\theta_\mathrm{j}$ \citep{lazzati2004jet, Farinelli_tophat}. However, theoretical models have advanced to considering jets with ``structure''. This structure is typically a jet with a fast moving core and a slower moving cocoon \cite{lazzati2018_GRB170817_afterglow,lazzati2019jet, salafia_EATS, lundman2018polarization}. 

The dynamics of the GRB jet can be understood under the simplistic fireball model \cite{cavallo1978fireball}. In this model energy is instantaneously released into a shell of high temperature plasma. The energy gets converted into photons at some distance close to the central engine, $r_0$. These photons then accelerate the material in the outflow to relativistic velocities. The ratio of the released energy to the rest mass energy of the material in the shell ($\eta=E/M c^2$) is an important factor in the dynamics of the fireball as it represents the ``baryon loading'' of the fireball\footnote{As the denominator increases, more baryons are included in the fireball which means that more energy is necessary to accelerate those baryons to a given bulk Lorentz factor. This bulk Lorentz factor is equivalent to $\eta$.}. {  Dynamically, this parameter describes different ways that fireballs can be radiation versus matter dominated. In the case of GRBs, we expect fireballs that are balanced between the two; the fireball should transfer its initial radiative energy to accelerating the baryons trapped in the outflow and then, when the matter is coasting at a relativistic speed, it should become optically thin to the radiation \citep{chhotray2018dynamic}. The baryon loading also defines different regimes in the evolution of a fireball, such as the radius where the fireball will stop accelerating and coast with a maximal bulk Lorentz factor equivalent to $\eta$ \citep[see e.g.][for a comprehensive overview]{peer_review}.} Thus, under the interpretation of GRB jets as fireballs, we are interested in placing constraints on $\eta$ (which is equivalent to constraining $\Gamma$ and the dynamics of the jet) and $r_0$.

There are many other physical quantities of interest in the jet, such as its composition (see e.g. \cite{Beloborodov_fuzzy_photosphere}), but these other quantities are not able to be probed using current observed correlations of GRB prompt emission. 

\section{Prompt Observed Correlations}\label{correlations}

\subsection{Luminosity-Time Scale Relations}
The peak isotropic luminosity-lag relation ($L_\mathrm{pk}-\tau_\mathrm{lag}$) was first explored by \citet{lumi_lag_relation_norris} who showed that low luminosity GRBs exhibit longer time lags than high luminosity bursts, meaning that the flux of low energy photons peak later than the flux of high energy photons for low luminosity bursts. This relation was found to be
\begin{equation}
    \frac{L_\mathrm{pk}}{10^{53}} \approx 1.3 \left(\frac{\tau_\mathrm{lag}}{0.01 \mathrm{~s}} \right)^{-1.14} ~{\rm erg~s}^{-1}
\end{equation}
which is in agreement with an independent analysis conducted by  \citet{lumi_lag_relation_schaefer}. Both analyses calculated the lag between BATSE channels 1 and 3 which correspond to energy ranges of 25-50 keV and 100-300 keV, respectively. Each analysis also calculated $L_{\rm pk}$ within 50-300 keV and the analysis conducted by  \citet{lumi_lag_relation_schaefer} uses the peak flux of the light curve binned into 256 ms time bins to calculate $L_{\rm pk}$\footnote{This analysis has been expanded more recently to understand how the spectral lag changes as a function of energy instead of just between 2 instrumental channels (see e.g. \cite{spectral_lag_shao}).}. 
This relationship was analyzed by  \citet{lumi_lag_ukwatta2010} using similar procedures for Swift BAT detected GRBs. They found similar correlations between $L_\mathrm{pk}$ and $\tau_\mathrm{lag}$ which was calculated using various combinations of instrument channels. For each combination of Swift BAT channels used to calculate the lag,  \citet{lumi_lag_ukwatta2010} provide best fit lines. A similar analysis was done by  \citet{lumi_lag_kawakubo2015} more recently to expand the sample of analyzed GRBs. The relationship that they obtain,
\begin{equation}
L_{\text {pk }}= 2.88^{+0.83}_{-0.65}  \times 10^{54} \left(\frac{\tilde{\tau}_\mathrm{lag}(1+z)^{-1}}{1 \mathrm{~s}}\right)^{-1.05 \pm 0.03} ~{\rm erg~s}^{-1}
\end{equation}
still has significant scatter. They calculate their $L_\mathrm{pk}$ by considering the 1 s peak photon flux in the 15–150 keV band and they calculate $\tau_\mathrm{lag}$ between energy bands of 15-25 keV and 50-100 keV.

\begin{figure}[t!]
\centering
\includegraphics[width=0.7\textwidth]{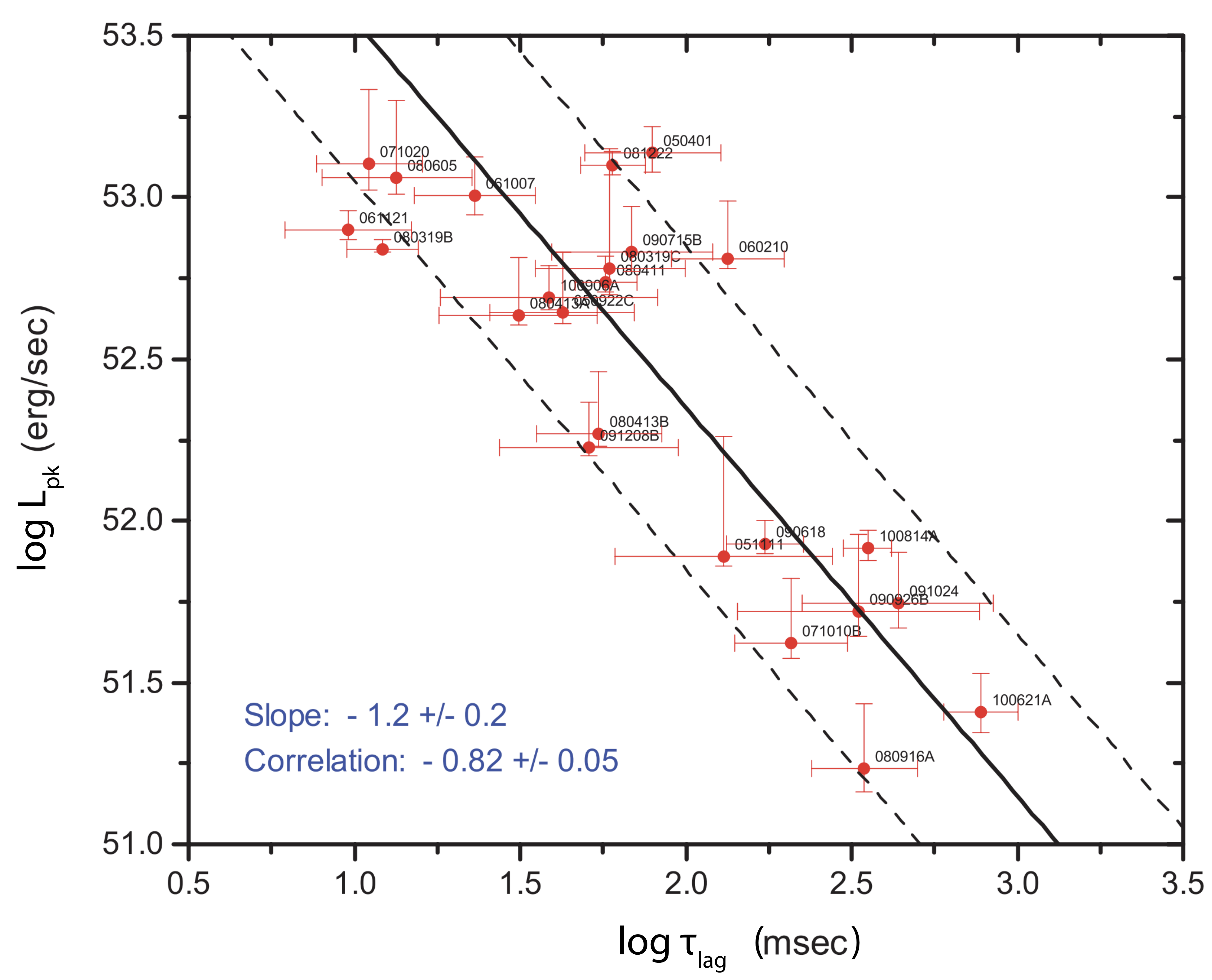}
\caption{The $L_{\rm pk}-\tau_\mathrm{lag}$ taken from \citet{lumi_lag_ukwatta2012}. The relationship given in Equation \ref{ukwatta_lumi_lag_rel}, is shown as the solid black line. The 1$\sigma$ confidence level is shown as the dashed lines and the GRBs used in their study are shown as the red markers with $1\sigma$ errors. The correlation coefficient for this relationship and the slope of the relationship are provided.
Credit: Ukwatta et al., The lag–luminosity relation in the GRB source frame: an investigation with Swift BAT bursts. Monthly Notices of the Royal Astronomical Society 2012, 419, 614–623. By permission of Oxford University Press on behalf of MNRAS. \copyright OUP
}
\label{ukwatta_lumi_lag}
\end{figure}

After,  \citet{lumi_lag_ukwatta2012} expanded on the study done by  \citet{lumi_lag_ukwatta2010} to provide constraints on the $L_{\rm pk}-\tau_\mathrm{lag}$ relation in the rest frame of GRBs. By considering the redshift of each GRB in their sample, they converted the observer channels to rest frame energies of 100–150 and 200–250 keV which were then used to calculate the lag. The $L_\mathrm{pk}$ was calculated by considering the 1 s peak photon flux in the 15–150 keV band. The relationship that  \citet{lumi_lag_ukwatta2012} obtained is:
\begin{equation}
    L_{\text {pk }}= 5.01^{+7.58}_{-3.02}  \times 10^{54}  \left(\frac{\tilde{\tau}_\mathrm{lag}(1+z)^{-1}}{0.001 \mathrm{~s}}\right)^{-1.2 \pm 0.2} ~{\rm erg~s}^{-1} \label{ukwatta_lumi_lag_rel}
\end{equation}
which is shown in Figure \ref{ukwatta_lumi_lag} as the solid black line and the estimated 1$\sigma$ confidence level as the dashed lines. 

There is an analogous relationship between the peak isotropic luminosity, $L_\mathrm{pk}$, and the rise time, $\tau_\mathrm{RT}$, for the detected GRB to reach its maximum flux, as shown by \citet{lumi_rt_schaefer}. This relation is given as
\begin{equation}
    L_\mathrm{pk} =  3.47^{+0.51}_{-0.45}  \times 10^{52}  \left( \frac{\tilde{\tau}_\mathrm{RT} (1+z)^{-1}}{0.1 \mathrm{~s}} \right)^{-1.21\pm 0.06} ~{\rm erg~s}^{-1}
\end{equation}
based on 69 BATSE and Swift detected GRBs that were analyzed. The $L_\mathrm{pk}$ is calculated by integrating over the fitted spectrum of detected photons normalized by the integrated spectrum in the energy range where the fitting was performed. $\tau_\mathrm{RT}$ is calculated by taking the shortest time over which the GRB light curve rises by 50\% of the peak flux, which involves smoothing the light curve. 

A slightly different relationship was also obtained by \citet{lumi_rt_xiao} by analyzing  107 GRBs. The $L_\mathrm{pk}-\tau_\mathrm{RT}$ relation that they obtain is
\begin{equation}
    L_\mathrm{pk} = 6.92^{+0.67}_{-0.61}  \times 10^{52} \left( \frac{\tilde{\tau}_\mathrm{RT} (1+z)^{-1}}{0.1 \mathrm{~s}} \right)^{ -1.70\pm 0.05} ~{\rm erg~s}^{-1}
\end{equation}
where the intercept and the slope of the relations are slightly different that what was found by  \citet{lumi_rt_schaefer}. \citet{lumi_rt_xiao} also revisited the $L_\mathrm{pk}-\tau_\mathrm{lag}$ relation and found an updated fit of
\begin{equation}
    L_\mathrm{pk} = 1.62^{+0.16}_{-0.14}  \times 10^{52} \left( \frac{\tilde{\tau}_\mathrm{lag} (1+z)^{-1}}{0.1 \mathrm{~s}} \right)^{- 0.98\pm 0.03} ~{\rm erg~s}^{-1}
\end{equation}

The $L_\mathrm{pk}-\tau_\mathrm{RT}$ relationship is closely tied to the relationship between $L_\mathrm{pk}$ and light curve variability, $V$, since the most important factor in determining light curve variability is the rise time of the light curve \cite{lumi_rt_schaefer}. This $L_\mathrm{pk}-V$ relation was first introduced by \citet{lumi_var_Reichart} who analyzed a sample of 20 GRBs. They obtained the general scaling 
\begin{equation}
    L_\mathrm{pk} \sim  V^{3.3^{+1.1}_{-0.9}} 
\end{equation}
Where the peak isotropic luminosity is measured from a 1 s binned light curve constructed from photons with energies between 50-300 keV. The variability is calculated by considering the difference between the detected light curve and a smoothed light curve. 

\citet{lumi_var_fenimore} developed a similar relation from their sample of 220 BATSE GRBs. The relation that they obtain is
\begin{equation}
    \frac{L_\mathrm{pk}}{d\Omega}=3.1\times 10^{56}V^{3.35} ~{\rm erg~s}^{-1} ~{\rm sr}^{-1}
\end{equation}
The variability is calculated in a similar methodology as \citet{lumi_var_Reichart}. The luminosity that \citet{lumi_var_fenimore} use here is calculated from the observed peak photon flux averaged over 256 ms between the 50-300 keV band. Additionally, the relation was constructed such that it can easily be modified if GRBs were collimated and not isotropic (which was not well established then), which is why the solid angle factor is explicitly placed in the relation.

\begin{figure}[h!]
\centering
\includegraphics[width=0.7\textwidth]{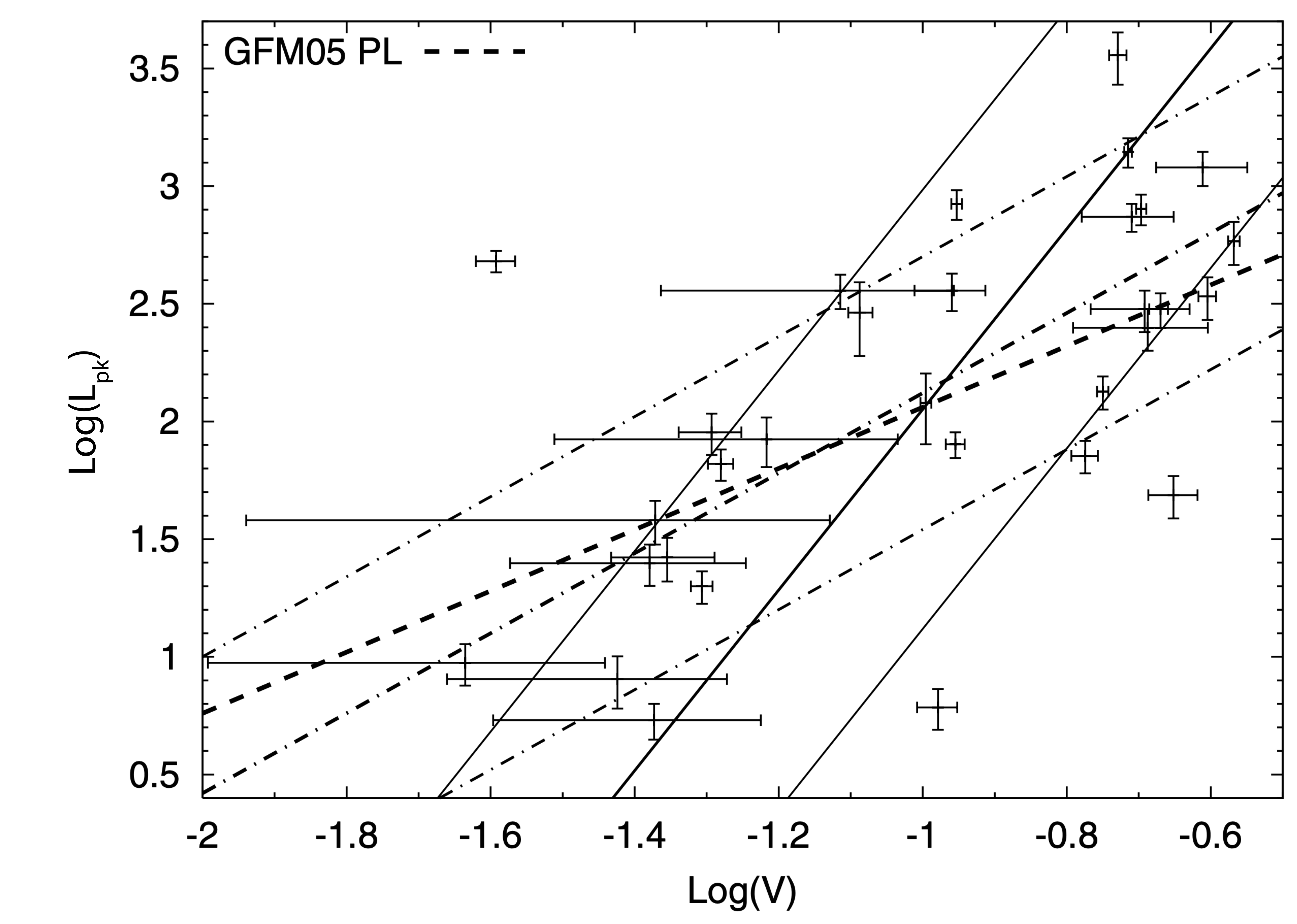}
\caption{The $L_\mathrm{pk}-V$ relation taken from \citet{lumi_var_slope_guidorzi} who used a number of different methods to calculate variability. Shown are the 32 GRBs with known redshifts that were analyzed by \citet{lumi_var_guidorzi}. The relationship that they obtained is plotted as the black dashed line. Also plotted are the $L_\mathrm{pk}-V$ relations obtained with two independent methods of calculating the slope of the relationship. The method outlined by \citet{lumi_var_guidorzi_reichart} and \citet{lumi_var_Reichart} is shown as the solid line with its $1\sigma$ confidence interval. The 
Bayesian method outlined by \citet{d2003bayesian} is shown as the dashed-dotted line with its $1\sigma$ confidence interval.
Credit: Guidorzi, et al., The slope of the gamma-ray burst variability/peak luminosity correlation. MNRAS, 2006, 371, 843–851. By permission of Oxford University Press on behalf of MNRAS. \copyright OUP}
\label{lumi_var}
\end{figure}

There are many other analyses of the $L_\mathrm{pk}-V$ relation, with differing sample sizes and smoothing methods (see e.g. \cite{lumi_var_guidorzi, lumi_var_guidorzi_reichart}). These studies have all recovered a relationship between $L_\mathrm{pk}$ and $V$ however the difference in the slope of each derived relation is relatively large when compared to other values. For example, \citet{lumi_var_slope_guidorzi} used two methods to obtain the slope of the $L_\mathrm{pk}-V$ relation and found slopes of $3.5^{+0.6}_{-0.4}$ and $0.88^{+0.12}_{-0.13}$, one of which is consistent with $3.3^{+1.1}_{-0.9}$ \cite{lumi_var_Reichart} while the other is very different, and in line with the slope obtained by \citet{lumi_var_guidorzi}. These discrepancies in the slope are shown in Figure \ref{lumi_var}, where various $L_\mathrm{pk}-V$ relations are plotted. Each relation has been obtained using different methods of calculating the slope of the relationship and thus show how unconstrained the relationship is. 

\subsubsection{Physical Interpretation}
While the $L_\mathrm{pk}-V$ relation is not well constrained, its existence can provide some insight into the outflow structure of GRB jets and the radiation processes that may take place within them. Under the internal shock model, the existence of this correlation that is observed can be contributed to shells of material colliding with one another in the GRB jet. Shells that are moving faster, with higher $\Gamma$, are more luminous and can produce highly variable light curves when they collide. This was shown by \citet{lumi_var_ramirez-ruiz_internal_shock} for a set of physical parameters. Under the assumptions that \citet{lumi_var_ramirez-ruiz_internal_shock} use, they find that:
\begin{itemize}
    \item Inverse Compton must dominate over synchrotron radiation in the emission region of the internal shock model, to produce spectral energies that are in agreement with highly variable GRBs \cite{lumi_var_lloyd_spectral}, and 
    \item Larger optical depths, $\tau>1$, to Compton scattering in the shells is necessary to: (1) produce lower energy photons for collisions that occur close to the central engine and (2) elongate temporal pulses of radiation that is produced in these collisions. 
\end{itemize}

The $L_\mathrm{pk}-V$ relation has also been interpreted under the photospheric model. \citet{lumi_var_kobayashi_internal_shock_photosphere} and \citet{lumi_var_meszaros_photosphere} who considered the effect of the location of the e$^\pm$ photosphere with respect to where the emission was being produced in the GRB jet. Their models still considered shells of material colliding within the jet as the primary source of radiation but the location of the photosphere changed the properties of the emitted radiation. If the shocks occurred below the photosphere, then the resulting emission was more diffuse and the pulse profiles were less variable\footnote{This type of physical setup can be classified as subphotospheric dissipation events. We denote these as photospheric models for this reason.}. If the shocks occurred above the photosphere, then the typical internal shock behavior is obtained. 

Under this general framework, \citet{lumi_var_kobayashi_internal_shock_photosphere} found the $L_\mathrm{pk}-V$ relation to be a result of a relationship between the jet opening angle, $\theta_\mathrm{j}$, and the variability of the light curve. This may be expected as the variability provides constraints on the emitting region of the GRB due to causality. Smaller emitting regions are a result of fast moving, more luminous jets with higher variability and smaller photospheric radii while larger emitting regions imply GRB jets with smaller $\Gamma$, lower luminosity, lower variability, and larger photospheric radii \cite{lumi_var_kobayashi_internal_shock_photosphere}. By simulating shells of material within GRB jets colliding with one another, \citet{lumi_var_kobayashi_internal_shock_photosphere} showed that differences in $\theta_\mathrm{j}$ from one GRB to the next can produce the $L_\mathrm{pk}-V$ relationship. 

The $L_\mathrm{pk}-V$ relationship under the photospheric model can also be due to the observer viewing angle changing with respect to the jet axis for each GRB, if one assumes a universal jet profile \cite{lumi_var_meszaros_photosphere, lumi_var_rossi_uni_jet, lumi_var_zhang_uni_jet, lumi_var_salmonson_uni_jet}.

Regardless of the emission process however, \citet{salafia_EATS} show that viewing angle effects are important to consider. The superposition of radiation from different regions of the jet combine such that a given light curve can become smoothed out and intrinsic variability is lost \cite{salafia_EATS}. 

The $L_\mathrm{pk}-\tau_\mathrm{RT}$ can be viewed as an extension of the $L_\mathrm{pk}-V$ relation. As \citet{lumi_rt_schaefer} point out, $\tau_\mathrm{RT}$ can be interpreted as the elapsed time from which the observer starts to receive radiation from the center of the emission region to when the entire emission region. (See \cite{salafia_EATS} for details on this process and its relation to the structure of light curves.) As we can infer from the $L_\mathrm{pk}-V$ relation, the size of the emitting region is dependent on $1/\Gamma$. Conversely, the luminosity of the emitting region scales proportionally with $\Gamma$. As a result, these two relationships combine to produce the  $L_\mathrm{pk}-\tau_\mathrm{RT}$ relation \cite{lumi_rt_schaefer}. 

The $L_\mathrm{pk}-\tau_\mathrm{lag}$ relation can be viewed as being closely related to the $L_\mathrm{pk}-V$ relation. As \citet{lumi_lag_zhang} highlight, variability should be inversely proportional to the spectral lag, with large variability jets exhibiting short spectral lags. Similarly to the $L_\mathrm{pk}-\tau_\mathrm{RT}$ relation, this is due to a large $\Gamma$ which produces smaller emission regions that the observer receives radiation from and thus the spectral lag from seeing photons from the center versus the edge of the region would be minimal \cite{lumi_rt_schaefer}. This interpretation is under the internal shock model. 

Using the internal shock model, \citet{lumi_lag_ioka} showed that the $L_\mathrm{pk}-\tau_\mathrm{lag}$ relation can also be caused by changes in observer viewing angle with respect to the jet axis. They find that bright GRBs are expected to have large $\Gamma$ and small observer viewing angles, meaning that the observer is viewing the GRB jet on axis. Correspondingly, the spectral lag should be small due to the smaller emitting region. Their findings are also relevant to the  $L_\mathrm{pk}-V$ relationship.

\citet{lumi_lag_uhm} used the framework of a single emitting shell that is producing synchrotron emission to explain the spectral lags. They showed that in order for this model to explain the data they need to enforce a number of physical constraints. Namely,
\begin{itemize}
    \item The radius of the emission region must be located $\gtrsim 10^{14} ~{\rm cm}$ away from the central engine
    \item The magnetic field strength within the emitting region must decrease as it expands
    \item There must be a burst of rapid bulk acceleration when the emission is produced
    \item The photon spectrum must have curvature associated with it (meaning it cannot be a single powerlaw across typical GRB energies)
    \item The spectral peak must progress from high to low energies as a function of time
\end{itemize}

Under the photospheric model, \citet{lumi_lag_jiang_photosphere} point out that a large portion of the physical constraints identified by \citet{lumi_lag_uhm} are fulfilled. In their study, \citet{lumi_lag_jiang_photosphere} show that the photospheric model can reproduce a general spectral lag evolution simply due to a change in optical depth to photons of different energies, as the jet expands. Radiative transfer simulations also show that structured jets can change the appearance of photospheric radiation and the time of arrival for photons of different energies \cite{parsotan_optical, Parsotan_spectropolarimetry}. 


\subsection{Spectral Peak Energy-Emitted Energy Relations}
Some of the first relationships that were explored related the spectral peak energy of observed GRBs with their measured energetics -- either the flux or the fluence of the burst. 

\citet{epk_flux_relation} found a positive correlation between the average peak energy of GRBs, $<\tilde{E}_\mathrm{pk}>$, and their peak fluxes, {  $\tilde{F}_\mathrm{pk}$}, both defined in the observer frame. They analyzed GRBs observed with BATSE and fit the observed GRB spectra with a cut-off power law in order to obtain the spectral peak energy. To obtain the peak fluxes of the GRBs, \citet{epk_flux_relation} binned the BATSE count rate data into 256 ms time bins for photons between 50 and 300 keV. The GRBs that were used in the correlation were selected such that {  $\tilde{F}_\mathrm{pk} > 1$} photon s$^{-1}$ cm$^{-2}$. Each GRB that survived the cut, was then sorted into five {  $\tilde{F}_\mathrm{pk}$} bins of varying widths. The bursts were sorted such that each bin progress from lowest {  $\tilde{F}_\mathrm{pk}$} values to the largest and the $\tilde{E}_\mathrm{pk}$ within each bin was averaged. This analysis produced the positive correlation between $<\tilde{E}_\mathrm{pk}>$ and {  $\tilde{F}_\mathrm{pk}$}. 

Subsequently, \citet{epk_fluence_relation} investigated the relationship between the average spectral peak energy of GRBs, $<\tilde{E}_\mathrm{pk}>$, and their fluence, $\tilde{S}_\mathrm{tot}$, both defined in the observer frame for the same energy range analyzed by \citet{epk_flux_relation}. 
The $\tilde{E}_\mathrm{pk}$ of each burst was measured by fitting a Band function to the observed spectrum.
The relationship that \citet{epk_fluence_relation} discovered is:
\begin{equation}
\tilde{E}_\mathrm{pk} \propto \tilde{S}_\mathrm{tot} ^{0.29\pm 0.03}
\end{equation}
\citet{epk_fluence_relation} also ensured that their results were consistent against instrumental biases and analyzed GRBs that were not near the detection threshold of BATSE. They found consistent results for $\tilde{E}_\mathrm{pk} \propto \tilde{S}_\mathrm{tot}$. 

Additionally, \citet{epk_fluence_relation} analyzed the $\tilde{E}_\mathrm{pk}-\tilde{F}_\mathrm{pk}$ relation for their populations of GRBs. \citet{epk_fluence_relation} correlated the flux with the $\tilde{E}_\mathrm{pk}$ measured in the time bin of $\tilde{F}_\mathrm{pk}$ and  found similar results to \citet{epk_flux_relation} for their entire sample of GRBs. However, when \citet{epk_fluence_relation} analyzed the $\tilde{E}_\mathrm{pk}-\tilde{F}_\mathrm{pk}$ relationship for the brightest GRBs in their sample, they found a weaker relationship compared to their whole sample.

When analyzing SGRBs, \citet{HIC_ghirlanda} found a similar relationship between $\tilde{E}_\mathrm{pk}$ and $\tilde{F}_\mathrm{pk}$.

\subsubsection{Physical Interpretation}
These relations between the spectral peak energy and the emitted energy of the GRB have led to the establishment of the Amati and Ghirlanda relations (see sections \ref{sec:amati_yonetoku} and \ref{sec:ghirlanda}). These relations, as we will see, relate the intrinsic radiated energy of the burst to the rest frame spectral peak energy.

Furthermore, the existence of $\tilde{E}_\mathrm{pk}-\tilde{F}_\mathrm{pk}$  relations for both LGRBs and SGRBs implies that the emission mechanism between both types of GRBs is the same \cite{HIC_ghirlanda}. 

\citet{brainerd1997cosmological} showed that the relationship discovered by \citet{epk_flux_relation} could not be the result of cosmological expansion of an otherwise static source. As a result this relationship should be something that is intrinsic to GRBs. \citet{epk_fluence_relation} analyze their obtained relationship in the context of the prompt emission of the GRB jet being due to internal versus external shocks that produce synchrotron radiation. They conclude that the $\tilde{E}_\mathrm{pk}-\tilde{S}_\mathrm{tot}$ is not compatible with the external shock model and is in alignment with what is expected from the internal shock model.

\subsection{Amati ($E_\mathrm{pk}$-$E_\mathrm{iso}$) and Yonetoku ($E_\mathrm{pk}$-$L_\mathrm{pk}$) Relations} \label{sec:amati_yonetoku}

\begin{figure}[ht]
\centering
\includegraphics[width=13cm]{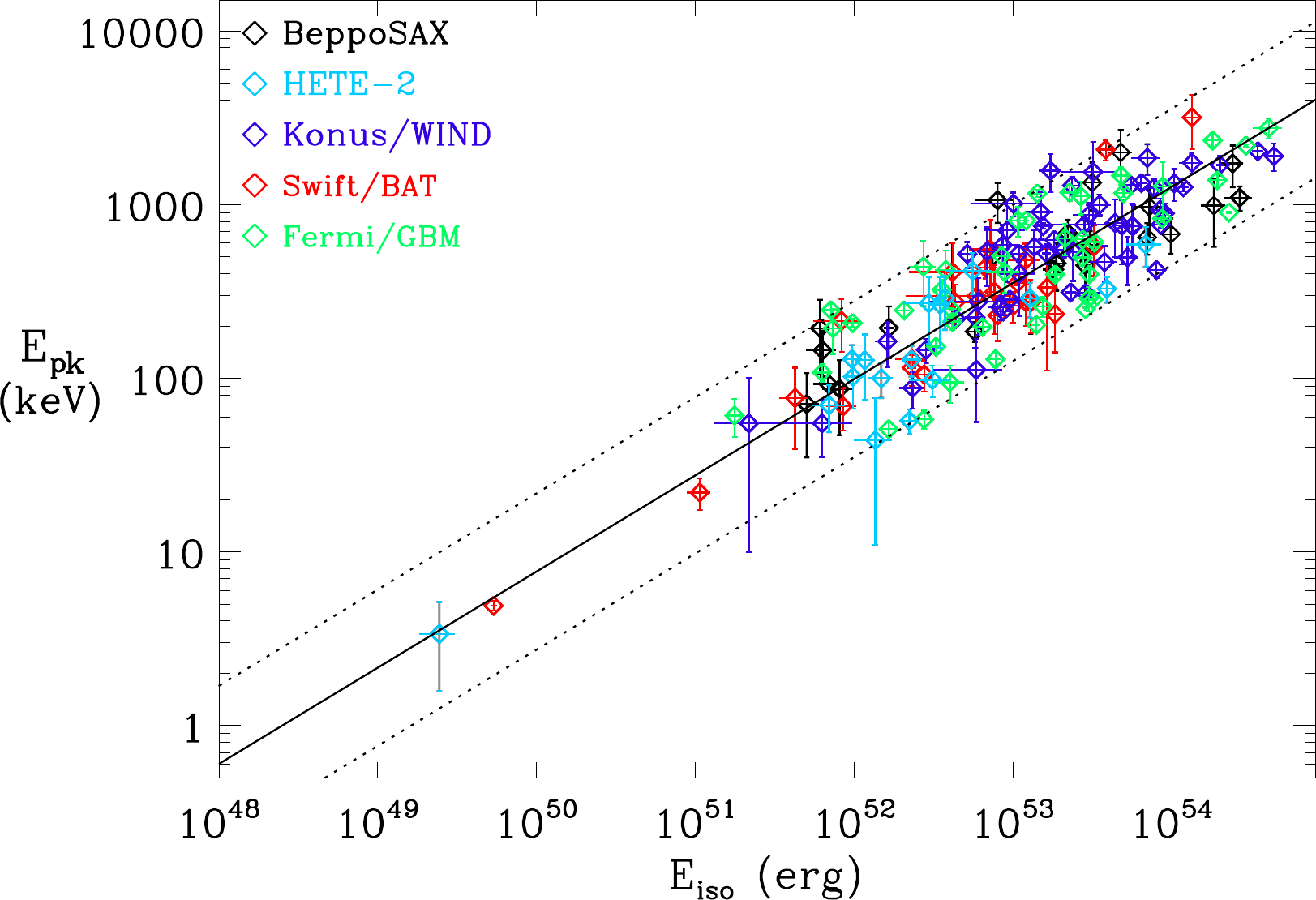}
\caption{The $E_\mathrm{pk}-E_\mathrm{iso}$ diagram of LGRBs taken from \citet{AV13}.  The solid and dashed lines show the best fit function and the 2$\sigma$ systematic errors of the Amati relation, respectively. Credit: Measuring Cosmological Parameters with Gamma Ray Bursts, Amati and Della Valle, International Journal of Modern Physics D, 22, 14, \copyright 2013 World Scientific
 }
\label{fig:Amati}
\end{figure}

One of the most studied empirical relations found in the prompt emission is the positive correlation between the peak energy of the time-integrated spectrum,  $E_\mathrm{pk}$, and the isotropic equivalent energy, $E_\mathrm{iso}$, which are defined in the GRB rest frame. The so-called Amati relation was initially discovered by \citet{Amati} based on the first sample of 12 LGRBs detected by BeppoSAX with redshift measurement (9 with firm redshift estimates and 3 with plausible values). The subsequent studies \citep[e.g.,][]{Amati06,data_set,AV13, Tsvetkova17} that explored larger number of samples detected by various instruments have confirmed and extended the correlation (Fig. \ref{fig:Amati}). The best fit power-law function to the correlation found by \citet{Zaninoni16} is
\begin{equation}
E_\mathrm{pk} = 110^{+150}_{-60}\left(\frac{E_\mathrm{iso}}{10^{52} ~{\rm erg}}\right)^{0.51\pm 0.04} ~{\rm keV}.
\label{Eq:Amati}
\end{equation}
The  uncertainties in the above formula indicate 2$\sigma$ confidence level.
As seen in Fig.\ref{fig:Amati}, the scatter of the correlation is not small ($\sigma \sim 0.2$ dex in $E_\mathrm{pk}$). While slightly different slopes are found for the correlation curve among the literature, they are consistent within the dispersion.

\begin{figure}[ht]
\centering
\includegraphics[width=13cm]{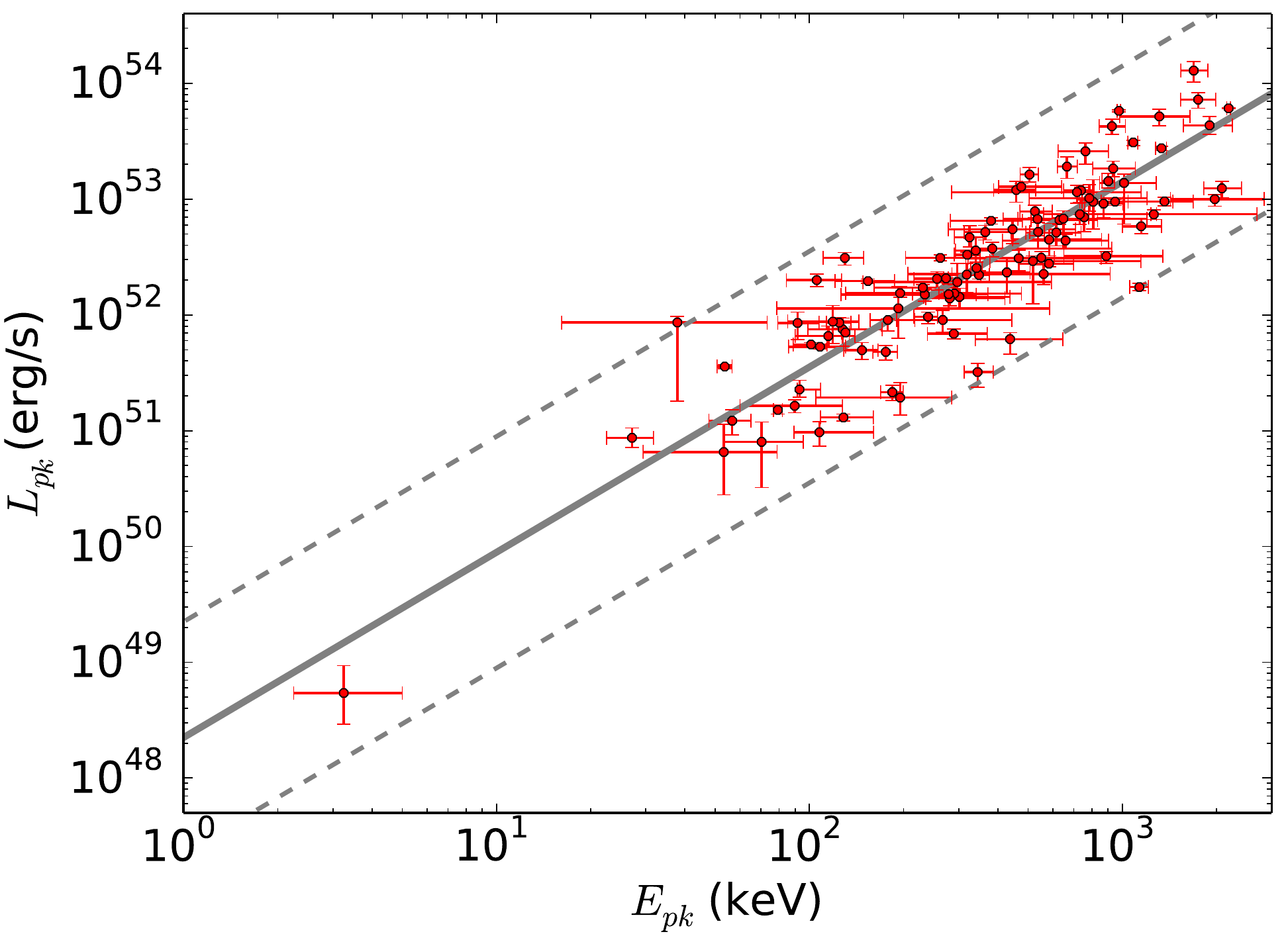}
\caption{The $E_\mathrm{pk}-L_\mathrm{pk}$ diagram of LGRBs taken from \citet{YMT10}. The solid and dashed lines show the best fit function and the 3$\sigma$ systematic errors of the Yonetoku relation, respectively. Data from  \citet{YMT10}.}
\label{fig:Yonetoku}
\end{figure}

Soon after the discovery of the Amati relation, \citet{Yonetoku} showed that there is a similar correlation between $E_\mathrm{pk}$ and the isotropic equivalent peak luminosity, $L_\mathrm{pk}$, using the BeppoSAX GRBs reported by \citet{Amati} with additional samples detected by BATSE (16 with firm redshift measurement and 3 with plausible values in total).  Intuitively, it is not surprising to find such a correlation since $E_\mathrm{pk}$ is likely to reflect the instantaneous physical condition of the emission  rather than time-integrated property $E_\mathrm{iso}$ (see Section \ref{sec:Golenetskii}). Subsequent studies also confirm the correlation with larger samples \citep[e.g.,][]{YMT10,data_set, Tsvetkova17} (Fig. \ref{fig:Yonetoku}). \citet{YMT10} finds a best-fit power-law function to the correlation given as
\begin{equation}
L_\mathrm{pk} = 2.7^{+8.5}_{-2.1}\times 10^{52} \left(\frac{E_\mathrm{pk}}{355~{\rm keV}}\right)^{1.60\pm 0.082} ~{\rm erg~s}^{-1},
\label{Eq:Yonetoku}
\end{equation}
where the uncertainties indicate 2$\sigma$ errors. 
The correlation seems to be slightly tighter than the Amati relation but still has non-negligible dispersion (According to the analysis of \cite{YMT10}, $\sigma = 0.33$ dex is found in $L_\mathrm{pk}$ in the Yonetoku relation, while $\sigma = 0.37$ dex is found in $E_\mathrm{iso}$ in the Amati relation). Slightly different slopes found in the literature are consistent within the dispersion. 

Several authors have pointed out the effect of observational biases that may artificially generate these correlations \citep{Nakar05,Band05,Butler07,Kocevski12}. However, careful analyses suggest that, while these effects may partially contribute, they cannot fully be responsible for producing the correlation \citep{Bosnjak08,Ghirlanda08,Nava11,data_set,Heussaff13}
 (see \cite{Dainotti18} for comprehensive review on the possible selection effects and systematic). Further evidence for the physical origin of the correlations comes from the fact that a similar correlation holds between $E_\mathrm{pk}$ and luminosity, $L_\mathrm{iso}$, measured in the time-resolved analysis of individual bursts \citep{Golenetskii,Firmani09,Ohno09,Ghirlanda10,ghirlanda2011sgrb_lgrb_same_emission_mech, Frontera12} (see Section \ref{sec:Golenetskii}). Therefore, it is likely that Amati and Yonetoku relations are intrinsic, while the scatter may be large.

These correlations span a wide range in $E_\mathrm{pk}$-$E_\mathrm{iso}$ and $E_\mathrm{pk}$-$L_\mathrm{pk}$ diagrams. They cover not only the ordinary LGRBs ($E_\mathrm{pk} \gtrsim 100~ {\rm keV}$) but also the X-ray-rich GRBs (XRRs) and X-ray flashes (XRFs) which are considered to be a continuous family (soft version) of LGRBs \citep{Sakamoto05, Lamb05, Sakamoto08}.
Interestingly, it is also found that at least some ULGRBs are consistent with these correlations \citep{Gendre13,Peng13,Zaninoni16}.\footnote{For example, the analysis by \citet{Zaninoni16} showed GRB 111209A, 101225A, and 130925A agree with the Amati relation.
The analysis of GRB121027A by \citet{Peng13} finds that, while  both the initial hard  emission and later soft emissions agree with the Yonetoku relation, Amati relation  is only satisfied for the initial emission.}
 On the other hand, it should also be noted that outliers to these correlations exist \citep{Nava08}\footnote{Outliers are often defined as a population located more than $\sim 2$-$3\sigma$ away from the correlation curve. Note that outliers are discarded in Figs. \ref{fig:Amati} and \ref{fig:Yonetoku}.},  although the observed number is subdominant. 
Notable examples are so-called low-luminosity GRBs (LLGRBs) \cite{D'Elia18}\footnote{LLGRBs are likely to be an outliers for both relations. An exception is GRB060218 which shows an agreement with Amati relation. The event is, however, a clear outlier of Yonetoku relation \cite{YMT10}.}, a sub-class of LGRBs that exhibit substantially lower luminosities. They do not follow the extrapolation from the ordinary LGRB luminosity function and are likely to dominate the local GRB population, suggesting a distinct origin for these events \citep{Virgili09, Sun15}.

\begin{figure}[ht]
\centering
\includegraphics[width=13cm]{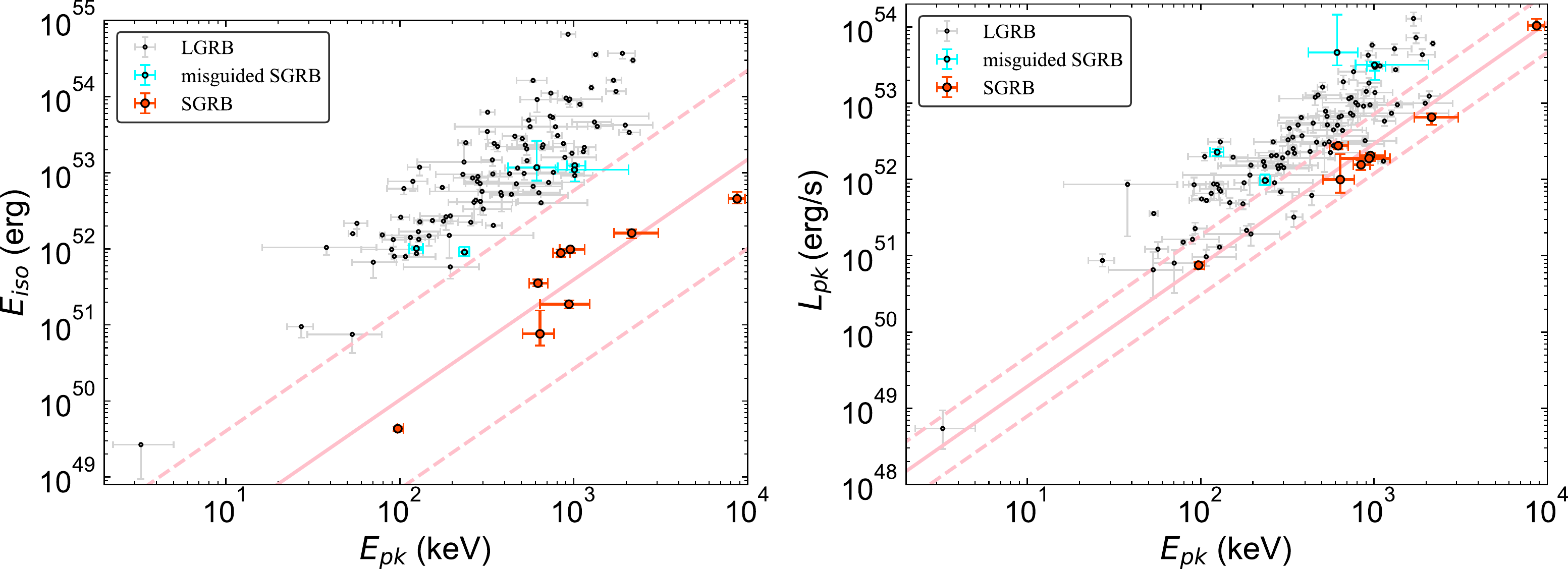}
\caption{The  $E_\mathrm{pk}$-$E_\mathrm{iso}$ ({\it left}) and $E_\mathrm{pk}$-$L_\mathrm{pk}$ ({\it right}) diagrams of SGRBs taken from \citet{Tsutsui13}.
According to the analysis of \citet{Tsutsui13}, red circles represent secure SGRBs, while cyan circles are considered
as misguided SGRBs. The best-fitting function and 3$\sigma$ intervals of the Amati ($E_\mathrm{pk}$–$E_\mathrm{iso}$) and Yonetoku ($E_\mathrm{pk}$–$L_\mathrm{pk}$) relations for the secure SGRBs are indicated with red
solid and dashed lines, respectively. 
For comparison, observational data of LGRBs taken from \citet{YMT10} are shown by gray points. Data from \citet{Tsutsui13} and \citet{YMT10}.
}
\label{fig:SGRB}
\end{figure}

Several authors have tested the existence of $E_\mathrm{pk}$-$E_\mathrm{iso}$ and $E_\mathrm{pk}$-$L_\mathrm{pk}$ correlations in SGRBs \citep{Amati06, Ghirlanda09, Zhang12, Tsutsui13,D'Avanzo14,Zaninoni16}. These studies suggest similar correlations also hold in SGRBs, although not as robust as in the case of LGRBs due to the limited number of events with redshift measurement. According to the analysis of \citet{Tsutsui13} based on a sample of 13 SGRB candidates (out of which 5 are considered to be falsely classified to SGRBs and therefore discarded), the correlations have a following form:
\begin{equation}
    E_\mathrm{iso} = 2.6^{+1.08}_{-7.68}\times 10^{51} \left(\frac{E_\mathrm{pk}}{774.5~{\rm keV}}\right)^{1.58\pm 0.28} ~{\rm erg},
\label{Eq:Amati_SGRB}
\end{equation}
\begin{equation}
    L_\mathrm{pk} = 1.95^{+0.32}_{-0.27}\times 10^{52} \left(\frac{E_\mathrm{pk}}{774.5~{\rm keV}}\right)^{1.59\pm 0.11} ~{\rm erg~s}^{-1}.
\label{Eq:Yonetoku_SGRB}
\end{equation}
The estimated dispersions in $E_\mathrm{iso}$ and $L_\mathrm{pk}$ for the above correlations are $\sigma = 0.39$ and $0.13$ dex, respectively (Fig. \ref{fig:SGRB}). 
As seen in the figure, SGRBs constitute a track parallel to LGRBs that are dimmer by a factor of $\sim 100$ ($\sim 5$) in $E_\mathrm{pk}$-$E_\mathrm{iso}$ ($E_\mathrm{pk}$-$L_\mathrm{pk}$) diagram.
While there are somewhat difference among the literature, independent studies
also seem to suggest that SGRBs lie on a much dimmer (slightly dimmer or consistent) track in the $E_\mathrm{pk}$-$E_\mathrm{iso}$ ($E_\mathrm{pk}$-$L_\mathrm{pk}$) diagram.

It should be noted that there is an ambiguity in selecting “true” SGRBs only from their duration, since LGRBs and SGRBs have an overlap in the duration distribution. Hence, it is essential to remove the possible contamination from LGRBs with a short duration to robustly constrain the properties of SGRBs.\footnote{\citet{Tsutsui13} attempt to discard "misguided" SGRBs by removing data that lies more than 3$\sigma$ away from the Amati relation of SGRBs.}
For example, recently found peculiar GRB200826A, which showed a rest frame duration of $\sim 0.5 s$, was confirmed to be association with a supernova \citep{Ahumada21}. Therefore, the event should be classified to the population of LGRBs despite its short duration. Indeed, GRB200826A is found consistent with the Amati and Yonetoku relations of LGRBs \cite{SvinkinGCN}.
It should also be noted that GRB170817A, associated with the LIGO-Virgo GW170817 binary neutron star merger event, is a clear outlier of the $E_\mathrm{pk}$-$E_\mathrm{iso}$ and $E_\mathrm{pk}$-$L_\mathrm{pk}$ correlations \citep{Hao17, Pozanenko18, Zhang18}. This indicates that, while GRB170817A provides the firm evidence for SGRBs to arise from the compact binary merger, its highly off-axis viewing angle results in a prompt emission property different from that of the ordinary SGRBs \citep[see e.g.,][]{Nakar20}.

\subsubsection{Physical Interpretation}
\label{sec:amati_yonetoku2}

Despite the considerable amount of theoretical works done up to now on the nature of GRB emissions, the origin of the prompt GRB correlations remains to be an open issue. The main difficulty comes from the fact that theoretical models do not have sufficient predictive power. This is particularly severe for synchrotron models in which the models themselves do not provide firm predictions on when and how much kinetic energy should be dissipated in the outflow. Moreover, there is an additional uncertainty on how the dissipated energy is transferred to the accelerated particles and magnetic fields. All these details are treated as model parameters, making it difficult to make quantitative estimations on $E_\mathrm{pk}$, $E_\mathrm{iso}$, and $L_\mathrm{pk}$. Therefore, to reproduce the observed correlations due to the diversity in the intrinsic nature of individual GRBs, one must assume that there is a self-regulation among imposed parameters \citep{Zhang_Meszaros02, ICMART_Zhang_2010, mochkovitch2015_amati_synch}. To our knowledge, there is no convincing explanation to satisfy such constraints.

Apart from the uncertainties on the dissipative processes, several studies have indicated that the viewing angle effect may be playing important role in shaping the observed correlations.
The most simplified case of a top-hat jet (a uniform outflow with a constant bulk Lorentz factor, $\Gamma$, and emissivity within a half-opening angle, $\theta_\mathrm{j}$) was explored by \citet{Yamazaki04}. Assuming that the Amati relation is satisfied for the on-jet emission (i.e., $\theta_\mathrm{v} < \theta_\mathrm{j}$, where $\theta_\mathrm{v}$ is the viewing angle of the observer), they demonstrated that the off-jet emission that shows a rapid decline in $E_\mathrm{pk}$ and $E_\mathrm{iso}$ with $\theta_\mathrm{v}$ also agrees with the relation up to $\theta_\mathrm{v} \sim 2\theta_\mathrm{j}$. 
Note that this viewing angle dependence is different from the well-known scaling for the point-source approximation (infinitesimally narrow jet: $\theta_\mathrm{j} \ll \theta_\mathrm{v}$), which exhibits $E_\mathrm{pk} \propto \delta(\theta_\mathrm{v})$ and $E_\mathrm{iso} \propto \delta(\theta_\mathrm{v})^3$, and hence $E_\mathrm{pk} \propto E_\mathrm{iso}^{1/3}$. 
Here $\delta(\theta_\mathrm{v}) = [\Gamma (1-\beta{\rm cos}\theta_\mathrm{v})]^{-1} $ is the Doppler factor, where $\beta$ is the velocity normalized by the speed of light.
While the point-source approximation is valid at $\theta_\mathrm{v} \gtrsim 2 \theta_\mathrm{j}$, the viewing angle dependence in the range $\theta_\mathrm{j} \lesssim \theta_\mathrm{v} \lesssim 2\theta_\mathrm{j}$ can be approximated as $E_\mathrm{pk} \propto \hat{\delta}(\theta_\mathrm{v})$ and $E_\mathrm{iso} \propto \hat{\delta}(\theta_\mathrm{v})^2$, where $\hat{\delta}(\theta_\mathrm{v})= [\Gamma \{1-\beta{\rm cos}(\theta_\mathrm{v} - \theta_\mathrm{j})\}]^{-1}$ \cite{Ioka18}.\footnote{Note that it is not straightforward to apply a similar estimation to the isotropic luminosity $L_{\rm iso}$ even for the case of uniform top-hat jet. This is because the luminosity depends not only on the viewing angle but also on the time evolution of the emissivity.} As a result, the effect of finite half-opening angle leads to a scaling $E_\mathrm{pk}\propto E_\mathrm{iso}^{1/2}$ which agrees with the Amati relation. A similar analysis was performed by \citet{Eichler04} and \citet{Levinson05} which showed that an annular-shaped jet can also reproduce Amati relation by the viewing angle effect. Subsequent studies have expanded the analysis to explore a more complicated jet structure \citep{Toma05, Graziani06, Donaghy06, Salafia15}. These works have shown that Amati correlation can be interpreted (at least partially) by the viewing angle effect
within the variety of jet models.\footnote{It should be noted that the formulations discussed here are only applicable only the optically thin emissions. Therefore, photospheric emission cannot be modeled.} 

Compared to the optically thin synchrotron models, photospheric models have a significant advantage in the predictive power since the emission is (quasi-)thermal. In this model, $E_\mathrm{pk}$ ($L_{\rm iso}$) is stabilized as it is fixed by the temperature (thermal energy) around the photosphere regardless of the dissipation mechanism. One of the consensuses in the recent studies is that the subphotospheric dissipation must take place at a large distance from the central engine to reproduce the observed correlations \citep{REES_MES_dissipative_photosphere, Thompson07}.\footnote{In the framework of the original fireball model \citep{goodman1986gamma, Paczynski86}, one naively expects a fireball to be initially formed at the vicinity of GRB central engine, $r_0 =10^6-10^7~{\rm cm}$. Hence, in the absence of subphotospheric dissipation, an adiabatic loss may be severe and reduce the values of $E_\mathrm{pk}$ and $L_{\rm iso}$ from those of the observations for a plausible baryon loading ($\eta \sim $ a few $100$).}
Various dissipative processes have been proposed up to now and explored in detail under the assumption of simplified spherical geometry \citep[e.g.,][]{Peer06, Giannios07, Beloborodov_fuzzy_photosphere, vurm_radiation}. These studies have indeed shown that photospheric emission can bring $E_\mathrm{pk}$ and $L_{\rm iso}$ in the range that agrees with the observations. Some authors \citep[e.g.,][]{Giannios12, Fan12, ito_stratified_jets} discuss possible interpretation for the origin of the correlation without specifying the details of the dissipation process. 
However, since the nature of the dissipative process is poorly understood, these arguments are far from conclusive.

Apart from the 1D models, studies of photospheric emission based on multi-dimensional hydrodynamical simulations have provided important progress on this issue. Initial works in 2D \citep{lazzati_variable_photosphere, Lazzati11, Mizuta11, Nagakura11} have shown that dissipation far above the central engine, a required condition mentioned above to reproduce typical values of $E_\mathrm{pk}$ and $L_{\rm iso}$, can be achieved by the collimation shock \citep[see also][for a sophisticated discussion]{Gottlieb19}. Soon after, \citet{lazzati_photopshere} and  \citet{diego_lazzati_variable_grb} suggested that the observed correlations can be interpreted as a sequence of viewing angles.
While these simulations have imposed a crude approximation on the emission process, their claim was verified by the later studies in which radiation transfer calculations are incorporated into hydrodynamical simulations in 2D \citep{MCRaT, parsotan_mcrat, parsotan_var, Ito21} and 3D \citep{Ito_3D_RHD, ito2019photospheric}. 

The $E_\mathrm{pk}$-$E_\mathrm{iso}$ and $E_\mathrm{pk}$-$L_\mathrm{pk}$ correlations found in these simulations reflect the angular structure of the hydrodynamic jet. At the central part of the jet up to a certain angle $\theta_{\rm core}$, there is a powerful “core” in which the power and Lorentz factor are high and close to uniform. Outside of the core region, an extended “wing” region is formed at the jet–cocoon interface, in which these quantities rapidly decrease with angle \citep[see e.g.,][for detail]{gottlieb2021structure}. As a result, for viewing angles that lie within the core ($\theta_\mathrm{v} \lesssim \theta_{\rm core}$), the emission properties, $E_\mathrm{pk}$, $L_\mathrm{pk}$, and $E_\mathrm{iso}$, do not vary largely with the viewing angle and exhibit values consistent with those of the typical GRBs. At larger viewing angles ($\theta_\mathrm{v} \gtrsim \theta_{\rm core}$), these quantities show a rapid decline, reflecting the sharp gradient in the wing region of the jet. This viewing angle dependence at the wing region reproduces a continuous sequence of $E_\mathrm{pk}$, $L_\mathrm{pk}$, and $E_\mathrm{iso}$ towards a dimmer population of GRBs.
These results indicate that Amati (Yonetoku) relation is a robust feature of the photospheric emission arising from a jet breaking out of a massive star or compact binary merger ejecta. 
A caveat to these simulation-based studies is that shock heating is the only dissipative process playing a role, and the various types of dissipative processes explored in the 1D models are not considered. Nevertheless, while further investigation is necessary, these studies provide a natural explanation for the origin of the observed correlations within a physically motivated setup.

\subsection{Golenetskii (Hardness-Intensity) Relation}
\label{sec:Golenetskii}
\citet{Golenetskii} proposed relating the luminosity (or flux) of a GRB to the effective temperature of the spectrum in each time bin of the light curve, thus providing constraints on the emission mechanism as a function of time. The temperatures of the spectra analyzed by \citet{Golenetskii} were obtained by fitting GRB time resolved spectra with a model consisting of a powerlaw with an exponential cutoff. That is: $F\propto E^{-\alpha}\exp(-E/E_0)$, where $\alpha$ is some powerlaw index and $E_0$ is the break energy that can be used to determine the temperature of the GRB plasma using $E_0=kT$. This model was motivated by the physics of thermal bremsstrahlung radiation being produced by an optically thin emission region of the jet that radiates at the energy $E_0$. In analyzing a number of Konus observed GRBs, \citet{Golenetskii} found a relationship that followed $L\propto (kT)^{\gamma}$ with $\gamma$ ranging from 1.5-1.7.

While the analysis was criticized by a number of studies \citep[see e.g.,][]{critique_HIC_laros, critique_HIC_norris1986, critique_recover_HIC_kargatis} the general correlation was recovered in many subsequent analyses \citep[see e.g.,][]{critique_recover_HIC_kargatis,HIC_borgonovo,HIC_guiriec,best_fit_lu,ghirlanda2011sgrb_lgrb_same_emission_mech, HIC_ghirlanda}. \citet{ghirlanda2011sgrb_lgrb_same_emission_mech} even showed in their analysis of $\sim 20$ Fermi GRBs that LGRBs and SGRBs follow the same correlation.

\begin{figure*}[t!]
\centering
\includegraphics[width=\textwidth]{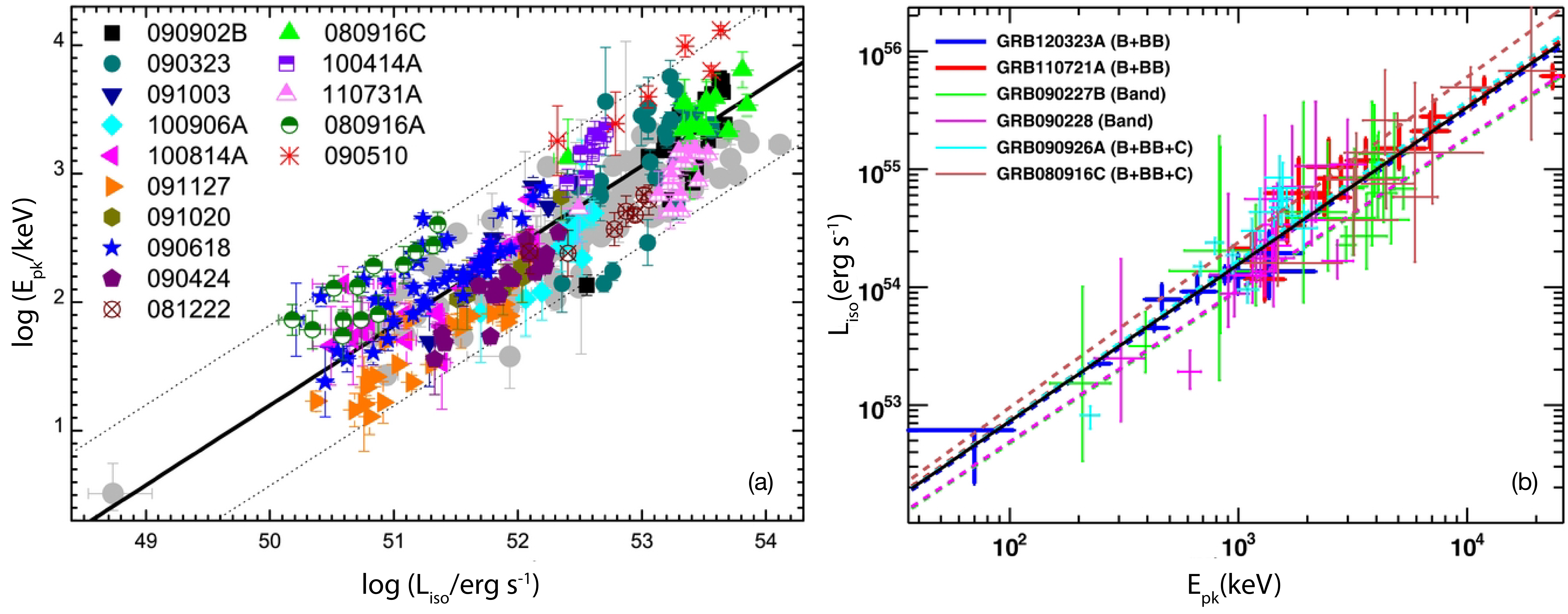}
\caption{The Golenetskii relations taken from \citet{best_fit_lu} (panel (a)) and \citet{HIC_guiriec} (panel (b)). In panel (a), the best fit given by Equation \ref{golenetskii_lu_eq} is shown as the solid black line and its $2\sigma$ dispersion is shown as the dotted lines. The 15 GRBs used to obtain the relation are shown as various marker types and colors. The gray circle markers are the GRBs from the sample used by \citet{YMT10} in the location that they would appear on the Yonetoku relation, thus showing that the dispersion in the Yonetoku relation is similar to the dispersion in the Golenetskii relation. Panel (b) shows the bet fit Golenetskii relation given in Equation \ref{golenetskii_guiriec_eq} as the solid black line. The different color points with errorbars show the location of various GRBs that were used to determine the best fit Golenetskii relation. The fits to individual GRBs are shown as dashed lines with identical colors. The multispectral components that are used to fit each bursts' spectra are denoted in the legend, where `B' or `Band' denotes Band, `C' denotes a cutoff power law, and `BB' denotes a blackbody spectrum. \copyright AAS. Reproduced with permission. }
\label{golenetskii_rel}
\end{figure*}

\citet{best_fit_lu} expanded on this finding to construct a universal relationship between isotropic luminosity, $\mathrm{L}_\mathrm{iso}$, calculated within 1–10$^4$ keV, and rest frame time resolved spectral peak energy $E_{\rm pk}$, which were obtained using Band function fits to the spectra. The relationship that they obtained is
 \begin{equation}
     \mathrm{L}_\mathrm{iso}= 1.17^{+2.72}_{-0.82}  \times 10^{48} \left(\frac{\mathrm{E}_\mathrm{pk}}{1 ~\mathrm{keV}}\right)^{1.61 \pm 0.01}  ~{\rm erg~s}^{-1} \label{golenetskii_lu_eq}
 \end{equation}
 and is shown in Figure \ref{golenetskii_rel}(a) as the solid black line and it's $2\sigma$ dispersion as the dotted lines. The obtained relationship was based on Fermi data of 14 LGRBs and 1 SGRB (GRB090510) with redshift measurements, which are also shown on the plot. \citet{best_fit_lu} also point out that their best-fit relationship has a similar scatter to that of the Yonetoku relationship.

This relation has been expanded upon by \citet{HIC_guiriec} who investigate the 
Golenetskii relation in observer quantities and in GRB rest-frame quantities for a small sample consisting of LGRBs and SGRBs. They find that the time resolved flux-observed spectral peak energy relation seems to be intrinsic within a given GRB but not among a sample of GRBs. Similar to \citet{best_fit_lu}, \citet{HIC_guiriec} find a relationship between the time resolved $\mathrm{L}_\mathrm{iso}$ and the time resolved rest-frame $\mathrm{E}_\mathrm{pk}$  which they report to be
\begin{equation}
    \mathrm{L}_\mathrm{iso}=(1.59\pm 0.84)\times 10^{50}\left(\frac{\mathrm{E}_\mathrm{pk}}{1 ~\mathrm{keV}}\right)^{1.33 \pm 0.07}  ~{\rm erg~s}^{-1} \label{golenetskii_guiriec_eq}
\end{equation}
This relation is shown in Figure \ref{golenetskii_rel}(b) as the solid black line. The GRBs that were used to obtain the relationship are shown as different colors and the similarly colored dashed lines denote the fit to that individual GRB. 
The analysis done by \citet{HIC_guiriec}
uses multispectral component models with the models denoted in the legend of Figure \ref{golenetskii_rel}(b). \citet{HIC_guiriec} point out that the multispectral component fits can provide better fits to GRB spectra than just the Band function and thus help reduce scatter in the Golenetskii relationship and other correlations presented in this review. 

The powerlaw slopes for the relationships obtained by \citet{best_fit_lu} and \citet{HIC_guiriec} are slightly different from one another which shows that the rest frame correlation is not well constrained. \citet{HIC_Burgess} used statistical methods to show that the current formulation of the Golenetskii relationship is not well constraining, at least in the context of using the relationship to detemine the redshifts of GRBs. They suggest adding terms to the relationship that can account for other physical parameters of GRBs such as the opening angle to make this relationship more universal among all GRBs. 

\subsubsection{Physical Interpretation}
\label{sec:Golenetskii2}
The Golenetskii relation \cite{Golenetskii} can be viewed as the time resolved Yonetoku relationship \cite{HIC_guiriec}, where the relationship is probing the prompt radiation mechanism in time. This time resolved relationship places important constraints on GRB models. 

The result that both SGRBs and LGRBS seem to follow this same correlation, implies that the radiation mechanism at play in both of these categories of GRBs is the same \cite{ghirlanda2011sgrb_lgrb_same_emission_mech, HIC_guiriec}. This places another constraint on prompt emission models by requiring the models to be independent of the progenitor. 

Theories predicting and explaining the Golenetskii relation are still in development \cite{HIC_Burgess, ghirlanda2011sgrb_lgrb_same_emission_mech} but there are a few results that have been obtained under each prompt emission model.

Under the internal shock model, \cite{HIC_bovsnjak_internal_shock} found a qualitative recovery of the Golenetskii relation but the slopes differed from those obtained from observational studies. Within their model, they find a better agreement with observations when they allow the fraction of accelerated electrons (which then cool and radiate photons) to be a function of the shock Lorentz factor. 

On the other hand, the photospheric model has been shown to be successful in reproducing the Golenetskii relation in LGRBs with little tuning \cite{diego_lazzati_variable_grb, parsotan_var}. These simulations of GRB jets consider a central engine that is turned on and off periodically. The stellar envelope, consisting of low energy plasma, is able to contaminate the jet during time periods when the central engine is turned off. As a result of this contamination, the average energy of electrons within the jet decreases which causes the optical depth to increase. As the optical depth increases photons increasingly scatter with the low energy electrons and gradually lose energy. Opposing this, portions of the jet that are ``turned on'' are not contaminated by the stellar envelope as much and maintain a relatively low optical depth. The fluid in this portion of the jet is relatively energetic which allows photons to maintain a large average energy when they scatter with the energetic electrons. The effect of the optical depth (on the cooling of the photons) is the main factor at play in these simulations and thus makes the results of these simulations applicable to situations when an observer is viewing a GRB far from the jet axis. The concept of the photospheric region, where the optical depth is different for different regions of the jet, is important in being able to naturally recover the Golenetskii relation \cite{parsotan_mcrat,parsotan_var}. 

\subsection{Ghirlanda Relation}
\label{sec:ghirlanda}
The Ghirlanda relation relates the source frame collimated corrected energy, $E_\gamma$, released by GRBs to their source frame spectral peak energy, $E_{\rm pk}$. The collimated corrected energy is calculated from the flux of the GRB, where the flux is calculated by using the best fit model spectrum to the time-integrated GRB spectrum and integrating the model spectrum from 1-10$^4$ keV. This model spectrum was also used to determine the time-integrated spectral peak energy of the GRB. The relationship for the 24 GRBs in their sample that have redshift measurements is found to be
\begin{equation}
    E_{\rm pk}  \sim 480  \left( \frac{E_\gamma}{10^{51} ~{\rm erg}} \right)^{0.7} {~\rm keV}
\end{equation}
which they point out has relatively small logarithmic scatter of $\sim 0.25$ dex. To obtain this result, \citet{ghirlanda_relation} had to make a number of assumptions including a constant density homogeneous circum-burst medium (CBM), used to calculate the jet opening angle from the afterglow measurements, a constant radiative efficiency, and a top-hat jet structure. 
 
\begin{figure*}[t!]
\centering
\includegraphics[width=\textwidth]{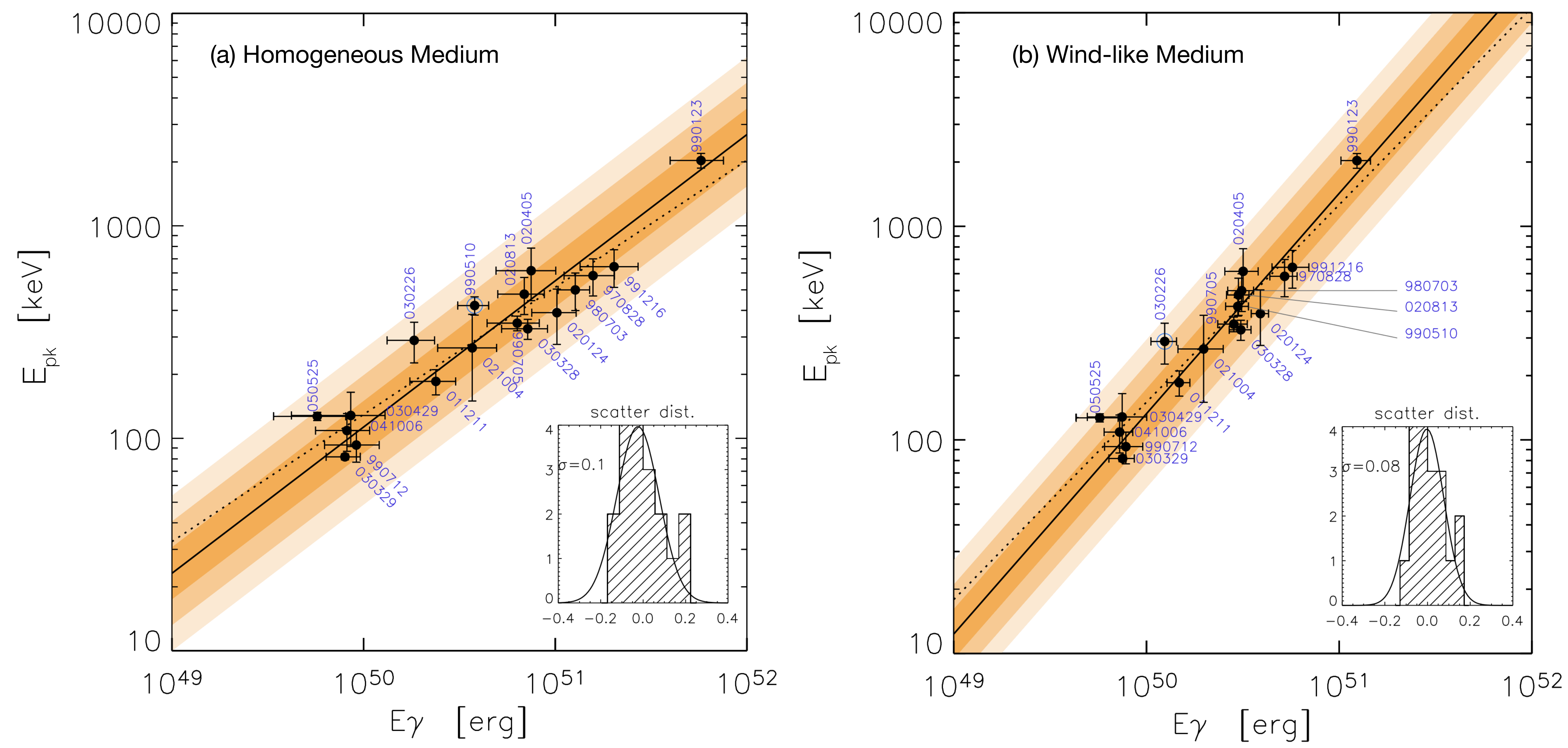}
\caption{The Ghirlanda relationship(s) taken from \citet{ghirlanda_nava}. In the panel (a) the relationship for the Ghirlanda relationship is shown assuming a homogeneous circumburst density. The yellow regions denote the 1$\sigma$, 2$\sigma$, and 3$\sigma$ scatter from darkest shading to lightest shading. The scatter is measured with respect to the best fit line, plotted as the solid black line (Equation \ref{ghirlanda_nava_eqhomo}), where uncertainties in both quantities were considered. The dotted line shows a simple regression analysis without considering the errors of each data point. The inset plot shows the scatter of the data with respect to the best fit line. The standard deviation of the best fit gaussian to the scatter distribution is also shown. Panel (b) is identical except the best fit relation is given by Equation \ref{ghirlanda_nava_eqwind}, for the assumption of a wind-like CBM. Credit: Nava et. al., A\&A, 450, 471-481, 2006, reproduced with permission \copyright ESO.}
\label{ghirlanda_nava_rel}
\end{figure*}

 The Ghirlanda relation was criticized for its use of these assumptions, with the greatest critique being based on its assumption of a homogeneous CBM. \citet{ghirlanda_nava} attempted to recover the Ghirlanda relation for a sample of 18 GRBs using two assumptions of the CBM that the afterglow is produced from: one for a homogeneous medium and another for a wind-like profile with a $r^{-2}$ dependence. The relationship that they find for the homogeneous CBM is:
 \begin{equation}
     E_{\mathrm{pk}}=(279 \pm 15)\left(\frac{E_{\gamma}}{2.72 \times 10^{50} \mathrm{erg}}\right)^{0.69 \pm 0.04} ~\mathrm{keV} \label{ghirlanda_nava_eqhomo}
 \end{equation}
while the relationship that they find under the wind-like medium is:
\begin{equation}
     E_{\mathrm{pk}}=(300 \pm 16)\left(\frac{E_{\gamma}}{2.2 \times 10^{50} \mathrm{erg}}\right)^{1.03 \pm 0.06} ~\mathrm{keV} \label{ghirlanda_nava_eqwind}
 \end{equation}
The reduced chi-squared values are 1.13 and 1.4, respectively, showing the tightness of the two correlations. These relations are shown in Figure \ref{ghirlanda_nava_rel}(a), under the homogeneous CBM assumption, and in Figure \ref{ghirlanda_nava_rel}(b), under the wind-like medium assumption.

\citet{ghirlanda_campana} then expanded on the study conducted by \citet{ghirlanda_nava} by analyzing a set of GRBs including a number of GRBs observed by Swift with redshifts and measured afterglow breaks. They found that the relations are still present, however the dispersion in the relations increase. They suggested that the increased scatter could be due to the more complex afterglow behaviors exhibited by Swift afterglows. 

Subsequently, \citet{ghirlanda_ghirlanda2007confirming} analyzed a set of 16 GRBs that fulfilled stringent requirements related to the determination of the jet break time, which factors into the calculation of the jet opening angle. Among other things, \citet{ghirlanda_ghirlanda2007confirming} required that the jet breaks be measured from the optical light curves of the afterglow which prevents unknown physics from affecting the measurements. The relationship that they obtained under the homogeneous CBM assumption is:
\begin{equation}
     E_{\mathrm{pk}}=(302 \pm 14)\left(\frac{E_{\gamma}}{4.4 \times 10^{50} \mathrm{erg}}\right)^{0.7 \pm 0.04} ~\mathrm{keV}
 \end{equation}
with a reduced chi-squared of 1.08 for 23 degrees of freedom.
The relation under the wind-like medium is:
\begin{equation}
     E_{\mathrm{pk}}=(334 \pm 18)\left(\frac{E_{\gamma}}{2.6 \times 10^{50} \mathrm{erg}}\right)^{1.05 \pm 0.06} ~\mathrm{keV}
 \end{equation}
with a reduced chi-squared of 0.89 for 23 degrees of freedom. \citet{ghirlanda_ghirlanda2007confirming} also highlight that the dispersions about the relationships are found to be gaussian with relatively small standard deviations ranging from $0.08-0.1$. 

Using a large sample of Swift GRBs, \citet{ghirlanda_wang2018gamma} revisited the Ghirlanda relation, using stringent selection criteria for the GRBs that they analyzed. As a result of their analysis they found a Ghirlanda relation similar to \citeauthor{ghirlanda_ghirlanda2007confirming}'s \cite{ghirlanda_ghirlanda2007confirming} homogeneous medium result, however  \citet{ghirlanda_wang2018gamma} find increased scatter in the relation. They attribute this scatter to the inclusion of GRBs with early time jet breaks in their sample.  

\subsubsection{Physical Interpretation}
The Ghirlanda relation, when it was initially proposed, did not have a well established interpretation as to how a geometric GRB property such as the collimated corrected energy (which is a function of the opening angle of the GRB jet) can affect a local emission process in the GRB jet, where the spectral peak energy is formed \cite{ghirlanda_relation, ghirlanda_liang}. 

\citet{ghirlanda_nava} showed that the Ghirlanda relation is Lorentz invariant, meaning that it still holds in the comoving frame of the GRB, and thus is intrinsic to all GRBS. Using this invariance, \citet{ghirlanda_nava} highlight that the number of photons in GRBs should be $\sim 10^{57}$ . This leads to a relationship where the energy that is injected into the GRB jet and the size of the injection region stay constant. An issue that arises here is that this result implies that the radiative efficiency of GRBs is variable, which is then in conflict with the use of the assumption of constant radiative efficiency which was necessary to obtain the Ghirlanda relation. This discrepancy may be resolved by considering that GRBs jets are structured and that we observe GRB jets at various observer viewing angles. 

\citet{ghirlanda2013faster} recently showed that the Ghirlanda, Amati, and Yonetoku relations are natural consequences of the jet opening angle being inversely related to the bulk Lorentz factor of the outflow. This relationship was also inferred based on the $L_{\rm pk}-V$ correlation as well and is seen in simulations of magnetically launched jets \cite{tchekhovskoy2009efficiency, komissarov2010rarefaction}.

To our knowledge, there have been no studies that have attempted to reproduce or interpret the Ghirlanda relation within the framework of the synchrotron or photospheric models.

\subsection{Lorentz Factor ($\Gamma_0$) - Prompt Emission ($E_\mathrm{iso}$, $L_{\rm pk}$, and $E_\mathrm{pk}$) Relations}
In the standard fireball model, the initial light curve of the forward shock afterglow is predicted to show a rising behavior (or maybe a plateau depending on the ambient density profile) during which the fireball is coasting with a constant bulk Lorentz factor, $\Gamma_0$ \citep{Rees92, Sari_1999_afterglow_reverse_shock, Kobayashi99, Kobayashi07}. The forward shock enters the deceleration phase once the energy of the swept-up matter becomes comparable to the initial energy of the fireball: $\Gamma_0^2 M_{\rm CBM} c^2 \sim E_\mathrm{k}$, where $M_{\rm CBM}$ and $E_\mathrm{k}$ are the swept-up mass of the CBM and the initial energy of the fireball\footnote{Assuming the efficiency of the prompt emission, $\epsilon$, the initial energy of the fireball in the afterglow phase can be estimated from $E_\mathrm{iso}$ as $E_\mathrm{k} = (1-\epsilon)\epsilon^{-1} E_\mathrm{iso}$.}, respectively. This transition marks a bump in the early afterglow, often termed “the onset of afterglow”, beyond which the light curve enters a decaying phase. Assuming such an interpretation, one can deduce the bulk Lorentz factor $\Gamma_0$ by measuring the peak time of early afterglow, $\tilde{t}_{\rm pk}$. 
Analytically, for a given energy $E_\mathrm{k}$ and a CBM density profile of $n_{\rm CBM}(r) = n_0 r^{-s}$, $\Gamma_0$ can be expressed as a function of $\tilde{t}_{\rm pk}$ as
\begin{equation}
\Gamma_0 = K \left(\frac{E_\mathrm{k}}{n_0 m_\mathrm{p} c^{5-s}}\right)^{\frac{1}{8-2s}} t_{\rm pk}^{-\frac{3-s}{8-2s}}
\label{Eq:Gamma0},
\end{equation}
where $t_{\rm pk}=\tilde{t}_{\rm pk}/(1+z)$ and $m_\mathrm{p}$ are the peak time in the rest frame and the proton mass, respectively.
Here $K$ is a numerical factor that depends on the power-law index of CBM density profile, $s$, and the underlying assumptions on the afterglow modeling, which is found to be within a factor of $\sim 3$ difference among the literature \citep[see][for detail]{Ghirlanda18}. 
As seen in the above equation, the dependence of $\Gamma_0$ on the parameters, $E_\mathrm{k}$ and $n_0$, is quite weak: $\Gamma_0 \propto (E_\mathrm{k}/n_0)^{1/8}$ in the case of the homogenous CBM ($s=0$), and $\Gamma_0 \propto (E_\mathrm{k}/n_0)^{1/4}$ in the case of the wind CBM ($s=2$).\footnote{In the case of the wind, the afterglow light curve does not necessarily show a rising phase before the deceleration (simple model predicts a rather flat behavior $\propto t^0$). However, as discussed in \citet{Ghirlanda12, Nappo14}, the initial rising phase may be realized by considering the pre-acceleration of the CBM.}
Hence, the estimation of $\Gamma_0$ from $t_{\rm pk}$ is not significantly affected by uncertainties in the model parameters, making it the more robust than the other proposed methods \citep[e.g., constraints from the high energy cut-off in the prompt emission;][]{Fenimore_grb_MeV_gamma_limit, Woods95, Lithwick01}.\footnote{It should be noted, however, that the method relies on the correct interpretation of early bump in the afterglow light curve to be associated with forward shock deceleration. Hence, if the bump has a different physical origin \citep[e.g., reverse shock;][]{Sari_1999_afterglow_reverse_shock, Zhang03}, it can lead to an erroneous conclusion. Also, possible ambiguities in the afterglow modeling (e.g., CBM density profile  \citep{Tian22},  dynamics of the forward shock deceleration \citep{Fukushima17}, and/or jet structure \citep{Racusin2008_grb080319B})  may also affect the accuracy of the estimation.}

The launch of the Swift satellite, together with the development of efficient follow-up by the ground-based telescopes, have enabled us to apply this method to a number of GRB afterglows in NIR, optical, and X-ray bands \citep{Molinari07, Xue09, Melandri10}. After the launch of Fermi, the detection of an early peak in the GeV light curve, if interpreted as an afterglow origin, has provided the earliest measurements of $t_{\rm pk}$ \citep{Ghisellini10, Kumar10, Ghirlanda10b}.
\citet{Liang10} performed a systematic study based on a sample of Swift GRBs in which 20 and 12 GRBs showed the onset feature in their optical and X-ray band, respectively. From the subsample of 17 optically selected GRBs with firm redshift measurements, they discovered a positive correlation between the bulk Lorentz factor and the prompt gamma-ray energy in the form $\Gamma_0 \propto E_\mathrm{iso}^{1/4}$ under the assumption of homogeneous CBM.

\begin{figure}[ht]
\centering
\includegraphics[width=13cm]{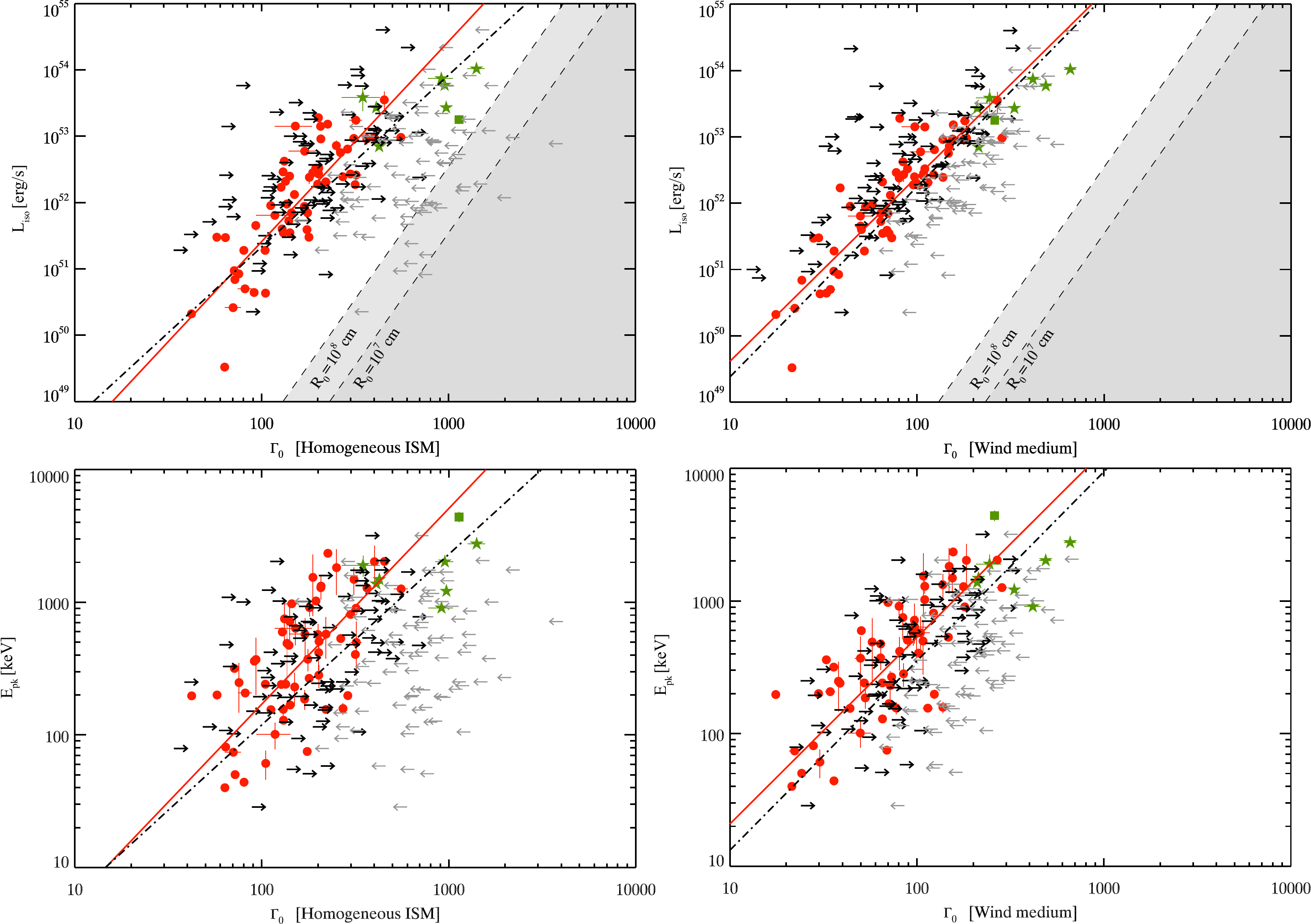}
\caption{The $\Gamma_0$-$L_\mathrm{pk}$ ({\it top}) and $\Gamma_0$-$E_\mathrm{pk}$ ({\it bottom}) correlations  taken from \citet{Ghirlanda18}. The left and right panels display the two cases in which homogenious and wind profile are assumed for the CBM, respectively. The red circles (green stars) correspond to the LGRBs with measured $t_{\rm pk}$ in optical (GeV) band, while the green square corresponds to short GRB090510. The rightward (leftward) arrows show the deduced values of lower (upper) limits on $\Gamma_0$ for GRBs without detection of $t_{\rm pk}$. Credit: Ghirlanda et. al., A\&A, 609, A112, 2018, reproduced with permission \copyright ESO.}
\label{fig:Gamma0}
\end{figure}

The existence of the $\Gamma_0$-$E_\mathrm{iso}$ correlation was verified by later studies, though somewhat difference is found in the best-fit slope \citep{Lu12, Ghirlanda12, Liang13}.
Moreover, the subsequent studies revealed that $L_\mathrm{pk}$ and $E_\mathrm{pk}$ are also positively correlated with $\Gamma_0$. Recently, \citet{Ghirlanda18} analyzed a large sample of GRBs composed of 67 with measured onset time $t_{\rm pk}$ in the optical or GeV light curves together with 106 with an upper limit. Of these, 66 with measured $t_{\rm pk}$ plus 85 with upper limits (and also deduced lower limits based on the assumption of $t_{\rm pk} > t_{90}$, where $t_{90}$ is the duration of the prompt emission) are used to investigate the correlations of $\Gamma_0$ with $E_\mathrm{iso}$, $L_\mathrm{pk}$ and $E_\mathrm{pk}$.\footnote{The inclusion of the sample with only upper limits in the analysis is important to avoid slection bias \citep[see][for detail]{Ghirlanda18}.}
As a result, they derived similar forms for the $E_\mathrm{iso}$-$\Gamma_0$ and $L_\mathrm{pk}$-$\Gamma_0$ correlations:
\begin{eqnarray}
    E_\mathrm{iso} & = &
    \begin{cases}
     0.955^{+0.425}_{-0.294}\times 10^{52} \left(\frac{\Gamma_0}{100}\right)^{2.19 \pm 0.10} ~{\rm erg} & {\rm for}~(s=0), \\
     1.48^{+0.0697}_{-0.0666}\times 10^{51} \left(\frac{\Gamma_0}{100}\right)^{2.66 \pm 0.05}~{\rm erg} & {\rm for}~ (s=2),
    \end{cases}\\
    L_\mathrm{pk} & = &
    \begin{cases}
     2.14^{+0.317}_{-0.276}\times 10^{51} \left(\frac{\Gamma_0}{100}\right)^{2.60 \pm 0.16} ~{\rm erg~s}^{-1} & {\rm for}~(s=0), \\
     5.62^{+0.265}_{-0.253}\times 10^{51} \left(\frac{\Gamma_0}{100}\right)^{2.87 \pm 0.05} ~{\rm erg~s}^{-1} & {\rm for}~ (s=2).
    \end{cases}
\end{eqnarray}
Here the upper (lower) part of each equation corresponds to the case of the homogeneous (wind) CBM. The $L_\mathrm{pk}$-$\Gamma_0$ correlation (shown in Fig. \ref{fig:Gamma0}) is found to be tighter than the $E_\mathrm{iso}$-$\Gamma_0$ correlation.
With a relatively large scatter, they obtained a correlation between $E_\mathrm{pk}$ and $\Gamma_0$ expressed as
\begin{eqnarray}
    E_\mathrm{pk}  = 
    \begin{cases}
     119^{+8.54}_{-7.97} \left(\frac{\Gamma_0}{100}\right)^{1.28 \pm 0.03} ~{\rm keV} & {\rm for}~(s=0), \\
     352^{+8.21}_{-8.02} \left(\frac{\Gamma_0}{100}\right)^{1.43 \pm 0.03}~{\rm keV} & {\rm for}~ (s=2).
    \end{cases}\\
\end{eqnarray}
All correlations show smaller dispersion in the case of wind CBM than in the case of homogenous CBM.

It is worth noting that, by combining these correlations, they derived relations $E_\mathrm{pk} \propto E_\mathrm{iso}^{0.58}$ ($E_\mathrm{pk} \propto E_\mathrm{iso}^{0.53}$) and $E_\mathrm{pk} \propto L_\mathrm{pk}^{0.50}$ ($E_\mathrm{pk} \propto L_\mathrm{pk}^{0.50}$) for the homogeneous (wind) CBM, which are consistent with the slopes of the Amati and Yonetoku relations. This result suggests $\Gamma_0$ may be a governing factor that determines the prompt emission properties, providing a possible hint for the origin of the Amati and Yonetoku relation. Moreover, as mentioned in section \ref{sec:ghirlanda}, $\Gamma_0$ may also be related to the origin of the Ghirlanda relation through the possible anticorrelation between $\Gamma_0$ and $\theta_{\rm j}$ \citep{ghirlanda2013faster}.

\subsubsection{Physical Interpretation}
The physical origin of the $\Gamma_0$-$E_\mathrm{iso}$, $\Gamma_0$-$L_\mathrm{pk}$, and $\Gamma_0$-$E_\mathrm{pk}$ correlations remains unclear. As mentioned in section \ref{sec:amati_yonetoku2}, the lack of predictive power in many theoretical models prevents us from making a detailed examination. Some studies attempted to provide possible interpretations without specifying the details of the emission process. For example, \citet{Liang10} suggested that correlations may be related to the angular structure of energy and Lorentz of the jet. Alternatively, \citet{Lu12, Lei13} proposed that these correlations may be attributed to the nature of GRB central engine properties.

In the context of the dissipative photospheric model, \citet{Giannios12, Fan12} suggested that the correlations can be reproduced within the reasonable range of model parameters.
On the other hand, the study by \citet{lazzati_photopshere} demonstrated that the photospheric emission evaluated based on the hydrodynamical simulations give rise to $\Gamma_0$ – $E_\mathrm{iso}$ correlation as a result of viewing angle effect. Although quantitative comparisons are yet to be made, subsequent series of simulation-based studies \citep{Ito_3D_RHD, MCRaT, parsotan_mcrat, parsotan_var, ito2019photospheric, Ito21} also seem to suggest that, together with the other prompt correlations (Amati, Yonetoku, and Golenetskii relations; see sections \ref{sec:amati_yonetoku2} and \ref{sec:Golenetskii2}), the correlations between the Lorentz factor ($\Gamma_0$) and prompt emission properties ($E_\mathrm{iso}$, $L_\mathrm{pk}$, and $E_\mathrm{pk}$) are a natural outcome for the photospheric emission from a hydrodynamic jet breaking out from a massive envelope.

\section{Conclusions}\label{conclusion}
We have outlined some of the most prominent prompt emission correlations that have shed light on the dynamics of GRB jets and the highest energy emission that is produced in these events. These correlations have probed a number of physical properties of GRBs such as their energy, timescales for emission, and spectral peak energies. 

The correlations shown here place important constraints on the plausible emission mechanisms that may produce the prompt emission in GRBs.  We have shown that the correlations can be explained under a number of different models. However, there still needs to be substantial work in connecting theoretical models to observational correlations. Conducting global radiative transfer calculations using GRB jets from hydrodynamic simulations has been a major step forward in bridging this divide but these types of analysis need to be conducted on a larger scale to produce predictions that can be compared to observations for a number of possible jet structures. Additionally, there is a need for full forward modeling of theoretically produced spectra and light curves through the responses of various instruments. This is done in other areas of astrophysics such as galaxy formation research \citep[see e.g.][]{cochrane2019predictions, parsotan2021realistic} where these methods have drastically pushed the field forward.

One important aspect of the observational correlations that we would like to emphasize is the use of different pieces of information of the GRB phenomenon. For example, using information from the afterglow can constrain geometric properties of GRB jets while the variability of the light curves can shed light on the `interior' structure of shells in the jets. These creative combinations of parameters have led to significant advances in our understanding. We expect that the next generation of GRB correlations will also take into account another highly sought after measurement of GRBs: polarization -- both time resolved and time integrated. With GRB polarization measurements becoming more precise, the combination of this measurement with some of the parameters that we have discussed in this review will significantly advance our understanding of GRB jet geometries and place even more stringent constraints on the radiation mechanism at play in these events. 

The future of GRB research is very bright partially due to their importance in multi-messenger astronomy and partially due to advanced instrumentation. As more advanced instrumentation and observational techniques are developed to measure GRBs, we need to take full advantage of the rich dataset that will be produced by advancing theoretical predictions of GRB prompt emission. Coupling global radiative transfer simulations with instrument responses is the primary way that we will be able to connect future observations to the dynamics and structure of GRB jets and provide strong predictions that can be tested against current and future GRB correlations. {  Multi-messenger detections of gravitational waves will also push our understanding of GRB prompt emission forward by providing independent constraints on the structure of GRB jets \citep{hayes2020_gw_sgrb_structure, beniamini2019lesson}, GRB jet launching mechanisms \citep{zhang2019delay}, which can guide prompt emission models, and it will help constrain the observer viewing angle of SGRB detections that are coincident with gravitational wave measurements, as in the case of GRB170817A \citep{grb_NS_merger_connection, lazzati2018_GRB170817_afterglow}. Additionally, detections of neutrinos coincident with GRB prompt emission measurements will place constraints on dissipation mechanisms that occur in GRB jets and constrain jets that fail to breakout of the progenitor star, thus providing insight into jet launching physics \citep[see ][ for a comprehensive review]{2022arXiv220206480K}.  }



\authorcontributions{writing---original draft preparation, T.P and H.I.; writing---review and editing, T.P and H.I. All authors have read and agreed to the published version of the manuscript.}

\funding{This research received no external funding.}

\acknowledgments{The material is based upon work supported by NASA under award number 80GSFC21M0002.  H.I. was supported by JSPS
KAKENHI grant (No. JP19K03878, JP19H00693, JP20H04751, and JP17H06362),  RIKEN
Interdisciplinary Theoretical \& Mathematical Science Program
(iTHEMS), and a RIKEN pioneering project “Evolution of Matter
in the Universe (r-EMU)” and “Extreme precisions to Explore
fundamental physics with Exotic particles (E3-Project)”.}

\conflictsofinterest{The authors declare no conflict of interest. The funders had no role in the writing of the manuscript or in the decision to publish this review.}

\reftitle{References}


\externalbibliography{yes}
\bibliography{references}

%

\end{document}